\documentstyle[preprint,aps,eqsecnum,12pt]{revtex}

\begin{document}
\draft
 
\preprint{\hfill\vbox{\baselineskip14pt
            \hbox{\bf KEK preprint 96-15}
            \hbox{\bf KEK-TH-456}
            \hbox{\bf SNUTP 95-111}}}
\title{\bf Probing CP Violation in $\gamma\gamma\rightarrow W^+W^-$\\
            with Polarized Photon Beams}
\author{S.Y.~Choi and K. Hagiwara}
\address{\it Theory Group, KEK, Tsukuba, Ibaraki 305, Japan }

\author{and\\
        M.S.~Baek}
\address{\it Department of Physics and Center for Theoretical Physics\\
             Seoul National University, Seoul 151-742, Korea}

\maketitle

\begin{center}
{Abstract}
\end{center}
\noindent
We demonstrate in a general framework that  
polarized photons by backscattered laser beams of adjustable 
frequencies at a TeV linear $e^+e^-$ collider provide 
us with a very efficient mechanism to probe CP violation
in two-photon collisions. CP violation in the process
$\gamma\gamma\rightarrow W^+W^-$ is investigated in detail 
with linearly polarized photon beams. There are two useful CP-odd 
asymmetries that do not require detailed information on $W$ decay 
products. The sensitivity to the CP-odd form factors are studied 
quantitatively by assuming a perfect $e$-$\gamma$ conversion and 
the 20 fb$^{-1}$ $e^+e^-$ integrated luminosity at the $e^+e^-$
c.m. energies $\sqrt{s}=0.5$ and $1.0$ TeV. 
The sensitivity is so high that such experiments will
allow us to probe new CP violation effects beyond the limits from 
some specific models with reasonable physics assumptions.
We find that a counting experiment of $W^+W^-$ 
events in the two-photon mode with adjustable laser frequencies 
can have much stronger sensitivity to the CP-odd $\gamma(\gamma) WW$ 
form factors than can a $W^+W^-$ decay-correlation experiment with a 
perfect detector achieve in the $e^+e^-$ mode.

\newpage
	
\section{Introduction}
\label{sec:introduction}

Even though the Standard Model (SM) has been successful in explaining 
all the experimental data up to date, it is believed that the SM is 
merely an effective theory valid at and below the weak scale and
that new physics beyond the SM should appear at higher energies.
We may expect to find new physics beyond the SM at high precision 
experiments on quantities whose SM values are suppressed. 
An interesting class of quantities where the SM contributions are strongly
suppressed are those with CP violation. In the SM, CP violation 
stems from the complex phase of the Kobayashi-Maskawa (KM) quark-mixing 
matrix\cite{MkMm} and the size of CP violation is often extremely small.
In contrast, various new physics scenarios on CP violation lead to 
comparatively large CP violation. Any CP-odd observable should hence be 
a good direct means to look for new physics effects. 

The next generation of $e^+e^-$ colliders\cite{NLC} will offer interesting
possibilities for studying physics of the heavy $H$, $t$-quark,
and the $W$ bosons either in the $e^+e^-$ mode or in the $\gamma\gamma$
mode. Linear collider physics in the $e^+e^-$ mode has been studied 
intensively for the past decade. Recently it has become clear that 
the $\gamma\gamma$ mode (as well as the $e\gamma$ mode)\cite{ScFs} can 
provide a good complement to experiments in the $e^+e^-$ mode.
For instance, it has been shown that the $\gamma\gamma$ mode has a unique 
advantage in the determination of the Higgs-two-photon coupling\cite{JgHh}
and its CP properties\cite{BgJg}. Pair production of the 
top-quark\cite{WbAb} and the $W$ boson\cite{JmBm,GbGc} in the $\gamma\gamma$
mode has also been studied as probes of CP violation in physics beyond
the SM. Most works\cite{WbAb,JmBm} have concentrated on the use of the spin 
correlations of the pair-produced top-quarks and the $W$ bosons which
require detailed study of their decay products. Recently, it has 
been pointed out\cite{ScKh} that the (linearly)-polarized photon beams can 
provide us with very powerful tests of the top-quark electric dipole moment 
(EDM) without any information on the $t\bar{t}$ decay patterns.  
Use of the $\gamma\gamma$ mode with linearly polarized photon beams for
studying CP violation in the process $\gamma\gamma\rightarrow W^+W^-$
has also been considered by B\'{e}langer and Coture\cite{GbGc}. 

In the present work we demonstrate in a rather general framework 
that polarized photons by backscattered laser beams of adjustable
frequencies provide us with a very efficient mechanism to probe
CP violation in the two-photon mode. 
We then give an extensive investigation of the possibility of probing 
CP violation with (linearly) polarized photon beams 
in the process $\gamma\gamma\rightarrow W^+W^-$ 
We extend in a systematic way the previous work\cite{GbGc}
so as to cover an arbitrary angle between polarization directions 
of two photon beams and an arbitrary laser beam frequency. 
We find in particular that adjusting of the laser beam frequency is 
essential to optimize the sensitivity to CP violation phenomena.
Furthermore, we study effects of all the possible dimension-six CP-odd 
operators composed of the Higgs doublet and the electroweak gauge bosons.

The $W^+W^-$ production in the $\gamma\gamma$ mode has 
several unique features in contrast to that in the $e^+e^-$ mode,
$e^+e^-\rightarrow W^+W^-$. In the $e^+e^-$ mode, a pair of $W$'s 
are produced via an annihilation of the colliding $e^-$ and $e^+$ where
the electronic chirality should be preserved along the electron 
line\cite{Kh} due to the very small electron mass in the SM. 
This forces the positron 
helicity to be opposite to the electron helicity such that the initial 
$e^+e^-$ configuration is always CP-even. 
On the contrary, there exists no apparent helicity selection mechanism
in the $\gamma\gamma$ production of $W^+W^-$.
This feature makes any CP-odd $\gamma\gamma$ configuration
in the initial state a good probe of CP-violation in the two-photon mode.

The process $\gamma\gamma\rightarrow W^+W^-$, which is 
characterized by the angle between the $W^+$ momentum and the $\gamma$ 
momentum in the c.m. frame, and the helicities of the particles, is C, P 
and CP self-conjugate, when the particle helicities are averaged over. 
For this reason the helicities (but not all of them) need to be determined 
or statistically analyzed to observe violations of these discrete 
symmetries.  One can take two approaches in analysing the process 
$\gamma\gamma\rightarrow W^+W^-$. One approach makes use of the spin
correlations of the two decaying $W$ bosons that can be measured by
studying  correlations in the $W^+W^-$ decay-product system, 
$(q\bar{q}^\prime)(l\bar{\nu})$ or $(\bar{l}\nu)(l\bar{\nu})$.  
The other method is to employ polarized photon 
beams to measure various polarization asymmetries of the initial states. 
Note that in the $e^+e^-$ mode, only the former method, the spin 
correlations of final decay products, is available. 
The two-photon mode allows 
us to combine the two methods. The use of the former technique in the 
two-photon collisions is essentially the same as that in $e^+e^-$ 
collisions\cite{KhRp,PkPm} with one crucial difference; in $e^+e^-$ 
collisions the spin of the $W^+W^-$ system is restricted to $J\geq 1$, 
while in $\gamma\gamma$ collisions $J=0$ is allowed. 
For a specific final state such as $W^+W^-$ and 
$t\bar{t}$ the two-photon cross section is larger than the corresponding 
$e^+e^-$ cross section. Especially, the $W$ pair cross section in the 
two-photon mode is much larger than that in the $e^+e^-$ mode because the 
$\gamma\gamma$ mode has contributions from the $J=0$ channel near 
threshold and a $t$-channel $W$ boson exchange. 
Moreover, it is easy to produce (linearly) polarized photon beams through 
the Compton back-scattering of polarized laser-light off the initial 
electron/positron beams. Hence the $\gamma\gamma$ mode of future linear 
colliders provides some unique opportunities to probe CP violation.

In Section~II we describe in a general framework how the photon 
polarization in the two-photon mode can be employed to study CP and 
CP$\tilde{\rm T}$ invariances.  Here {$\tilde{\rm T}$ is the so-called 
naive-time-reversal operation which reverses the signs of the 
three-momenta and spin of all particles but does not reverse the 
direction of the flow of time.  
The notation is introduced in Ref.~\cite{KhRp}.} 
Assuming that two-photon beams are purely linearly polarized in the 
colliding $\gamma\gamma$ c.m. frame we construct two CP-odd and 
CP$\tilde{\rm T}$-even asymmetries which allow us to probe CP violation 
without any direct information on the momenta and polarization of 
the final-state particles. All that we have to do is to count the 
number of signal events for a specific polarization 
configuration of the initial two photons. 
In Section~III we give a short review of a mechanism of producing 
highly energetic photons, the Compton backscattering of laser photons 
off the electron/positron beam\cite{GKS}, and we introduce two 
functions that measure the partial transfers of the linear 
polarization from the laser beams to the Compton back-scattered 
photon beams. We then investigate in detail which parameters are 
crucial to optimize the observality of CP violation with linearly 
polarized laser beams. 

In Section~IV we study consequences of CP violating new interactions 
in the bosonic sector of the SM, by adopting a model-independent 
approach where we allow all six dimension-six operators of the 
electroweak gauge bosons and the Higgs doublet that are 
CP-odd\cite{HaSt}. We identify all the vertices and present the 
Feynman rules relevant for the process $\gamma\gamma\rightarrow W^+W^-$. 
In Section~V, including all the new contributions, we present the 
helicity amplitudes of the $\gamma\gamma \rightarrow W^+W^-$ reaction. 
Folding with the effective two-photon energy 
spectrum, we then estimate the size of the two CP-odd asymmetries 
for a set of CP-odd operator coefficients. 
In Section~VI we present the $1$-$\sigma$ sensitivities 
to the CP-odd parameters by assuming a perfect $e$-$\gamma$ 
conversion in the Compton backscattering mechanism for an a $e^+e^-$ 
integrated luminosity of 20 fb$^{-1}$. 
We then compare the sensitivities in
the two-photon mode with those in the $e^+e^-$ mode under 
the same luminosity and c.m. energy, by restricting ourselves 
to the $W$ EDM and the $W$ magnetic quadrapole moment (MQD). 
Finally in Section~VII we summarize our findings and give conclusions.
\vskip 1.5cm

\section{Photon Polarization}
\label{sec:polarization}

In this section we fix our notation to describe in a general 
framework how photon polarization can provide us with an efficient 
mechanism to probe CP and CP$\tilde{\rm T}$ invariances
in the two-photon mode. With purely linearly-polarized 
photon beams, we classify all the distributions according to
their CP and CP$\tilde{\rm T}$ properties. 
Then, we show explicitly how linearly 
polarized photon beams allow us to construct two CP-odd and 
CP$\tilde{\rm T}$-even asymmetries which do not require  
detailed information on the momenta and polarization of the 
final-state particles. 

\subsection{Formalism}

A photon should be transversely polarized. 
For the photon momentum in the positive $z$ direction 
the helicity-$\pm 1$ polarization vectors are given by 
\begin{eqnarray}
|\pm >=\mp\frac{1}{\sqrt{2}}\left(0,1,\pm i,0\right).
\end{eqnarray}
Generally, a purely polarized photon beam state is a linear combination
of two helicity states and the photon polarization vector  
can be expressed in terms of two angles $\alpha$ and $\phi$
in a given coordinate system as
\begin{eqnarray}
|\alpha,\phi\rangle =-\cos(\alpha) e^{-i\phi}|+\rangle
           +\sin(\alpha) e^{i\phi}|-\rangle,
\end{eqnarray}
where $0\leq \alpha\leq \pi/2$ and $0\leq \phi\leq 2\pi$.
Then, the $2\times 2$ photon density matrix $\rho$\cite{GKS,VbEl} 
in the helicity basis $\{|+\rangle , |-\rangle \}$ is given by
\begin{eqnarray}
\rho\equiv |\alpha,\phi\rangle \langle \alpha,\phi |
    = \frac{1}{2}\left(\begin{array}{cc}
      1+\cos(2\alpha)\ \  & -\sin(2\alpha){\rm e}^{2i\phi}\\
           {  }           &      {  }                    \\
      -\sin(2\alpha){\rm e}^{-2i\phi}\ \ & 1-\cos(2\alpha)
      \end{array}\right).
\label{density matrix}
\end{eqnarray}
It is easy to read from Eq.~(\ref{density matrix}) that the degrees 
of circular and linear polarization are, respectively,  
\begin{eqnarray}
\xi=\cos(2\alpha),\qquad \eta=\sin(2\alpha),
\end{eqnarray}
and the direction of maximal linear polarization is denoted
by the azimuthal angle $\phi$ in the given coordinate system. 
Note that $\xi^2+\eta^2=1$ as expected
for a purely polarized photon. For a partially polarized photon
beam it is necessary to rescale $\xi$ and $\eta$ by its degree
of polarization $P$ ($0\leq P\leq 1$) as
\begin{eqnarray}
\xi=P\cos(2\alpha),\qquad \eta=P\sin(2\alpha),
\end{eqnarray}
such that $\xi^2+\eta^2=P^2$.

Let us now consider the two-photon system in the center-of-mass frame 
where one photon momentum is along the positive $z$ direction. 
The state vector of the two photons is 
\begin{eqnarray}
|\alpha_1,\phi_1;\alpha_2,\phi_2\rangle
 &=&|\alpha_1,\phi_1\rangle|\alpha_2,-\phi_2\rangle\nonumber\\ 
 &=&\cos(\alpha_1)\cos(\alpha_2)\: e^{-i(\phi_1-\phi_2)}|++\rangle
   -\cos(\alpha_1)\sin(\alpha_2)\: e^{-i(\phi_1+\phi_2)}|+-\rangle
    \nonumber\\
 & &-\sin(\alpha_1)\cos(\alpha_2)\: e^{i(\phi_1+\phi_2)}|-+\rangle
   +\sin(\alpha_1)\sin(\alpha_2)\: e^{i(\phi_1-\phi_2)}|--\rangle,
\label{two-photon_wf}
\end{eqnarray}
and then the transition amplitude from the polarized two-photon state 
to a final state $X$ in the two-photon c.m. frame is simply given by
\begin{eqnarray}
\langle X |M|\alpha_1,\phi_1;\alpha_2,\phi_2\rangle .
\end{eqnarray} 
The azimuthal angles $\phi_1$ and $\phi_2$ are the directions
of maximal linear polarization of the two photons, respectively, in 
a common coordinate system (see Fig.~1). 
In the process $\gamma\gamma\rightarrow W^+W^-$, 
the scattering plane is taken to be the $x$-$z$ plane
in the actual calculation of the helicity amplitudes. 
The maximal linear polarization angles are then chosen as follows. 
The angle $\phi_1$ ($\phi_2$) is the azimuthal angle of 
the maximal linear polarization of the photon beam, whose momentum 
is in the positive (negative) $z$ direction, with respect to 
the direction of the $W^+$ momentum. 
Note that we have used $|\alpha_2,-\phi_2\rangle$ in 
Eq.~(\ref{two-photon_wf}) for the photon whose momentum is along the 
negative $z$ direction in order to employ a common coordinate system 
for the two-photon system. 

For later convenience we introduce the abbreviation
\begin{eqnarray}
M_{\lambda_1\lambda_2}=\langle X|M|\lambda_1\lambda_2\rangle,
\end{eqnarray}
and two angular variables:
\begin{eqnarray}
\chi=\phi_1-\phi_2,\qquad 
\phi=\phi_1+\phi_2,
\end{eqnarray}
where $-2\pi\leq \chi\leq 2\pi$ and $0\leq \phi\leq 4\pi$ 
for a fixed $\chi$.
It should be noted that (i) the azimuthal angle difference, $\chi$, 
is independent of the final state, while the azimuthal angle sum, 
$\phi$, depends on the scattering plane, and (ii) both angles are 
invariant with respect to the Lorentz boost along the two-photon 
beam direction. 

It is now straightforward to obtain the angular dependence of the
$\gamma\gamma\rightarrow X$ cross section on the initial beam 
polarizations 
\begin{eqnarray}
\Sigma(\xi,\bar{\xi};\eta,\bar{\eta};\chi,\phi)
&\equiv& \sum_X|<X|M|\xi,\bar{\xi};\eta,\bar{\eta};\chi,\phi>|^2
 \nonumber\\
&=&\frac{1}{4}\sum_X
   \bigg[|M_{++}|^2+|M_{+-}|^2+|M_{-+}|^2+|M_{--}|^2\bigg]\nonumber\\
&& +\frac{\xi}{4}\sum_X
   \bigg[|M_{++}|^2+|M_{+-}|^2-|M_{-+}|^2-|M_{--}|^2\bigg]\nonumber\\
&& +\frac{\bar{\xi}}{4}\sum_X
   \bigg[|M_{++}|^2-|M_{+-}|^2+|M_{-+}|^2-|M_{--}|^2\bigg]\nonumber\\
&& +\frac{\xi\bar{\xi}}{4}\sum_X
   \bigg[|M_{++}|^2-|M_{+-}|^2-|M_{-+}|^2+|M_{--}|^2\bigg]\nonumber\\
&& -\frac{\eta}{2}Re\Bigg[{\rm e}^{-i(\chi+\phi)}
   \sum_X\bigg(M_{++}M^*_{-+}+M_{+-}M^*_{--}\bigg)\Bigg]\nonumber\\
&& -\frac{\bar{\eta}}{2}Re\Bigg[{\rm e}^{-i(\chi-\phi)}
   \sum_X\bigg(M_{++}M^*_{+-}+M_{-+}M^*_{--}\bigg)\Bigg]
   \nonumber\\
&& -\frac{\eta\bar{\xi}}{2}Re\Bigg[{\rm e}^{-i(\chi+\phi)}
   \sum_X\bigg(M_{++}M^*_{-+}-M_{+-}M^*_{--}\bigg)\Bigg]
   \nonumber\\
&& -\frac{\bar{\eta}\xi}{2}Re\Bigg[{\rm e}^{-i(\chi-\phi)}
   \sum_X\bigg(M_{++}M^*_{+-}-M_{-+}M^*_{--}\bigg)\Bigg]\nonumber\\
&& +\frac{\eta\bar{\eta}}{2}
   Re\Bigg[{\rm e}^{-2i\phi}\sum_X\bigg(M_{+-}M^*_{-+}\bigg)
          +{\rm e}^{-2i\chi}\sum_X\bigg(M_{++}M^*_{--}\bigg)\Bigg],
\label{distribution}
\end{eqnarray}
where the summation over $X$ is for the polarizations of the final states, 
and $(\xi, \bar{\xi})$ denote the degrees of circular polarization
and $(\eta,\bar{\eta})$ denote those of linear polarization of the 
two initial photon beams, respectively. They are expressed in terms of 
two parameters $\alpha_1$ and $\alpha_2$ by   
\begin{eqnarray}
&&\xi=P\cos(2\alpha_1),\qquad \bar{\xi}=\bar{P}\cos(2\alpha_2),
  \nonumber\\
&&\eta=P\sin(2\alpha_1),\qquad \bar{\eta}=\bar{P}\sin(2\alpha_2),
\end{eqnarray}
where $P$ and $\bar{P}$ ($0\leq P,\bar{P}\leq 1$) are the polarization 
degrees of the two colliding photons.

It is easy to check that there are sixteen independent terms, 
which are all measurable in polarized two-photon collisions. 
We find that purely linearly polarized photon beams allow us to 
determine nine terms among all the sixteen terms, while
purely circularly polarized photon beams allow us to determine
only four terms. The first term in Eq.~(\ref{distribution}), which 
corresponds to the unpolarized cross section, is determined in 
both cases. However, both circular and linear polarizations are 
needed to determine the remaining four terms. 

Even though we obtain more information with both circularly and 
linearly polarized beams, we study in this paper mainly the case 
where two photons are linearly polarized but not circularly polarized.
The expression of the angular dependence then greatly simplifies to
\setcounter{equation}{0}
\renewcommand{\theequation}{\arabic{section}.12\alph{equation}}
\begin{eqnarray}
{\cal D}(\eta,\bar{\eta};\chi,\phi)
&=&\Sigma_{\rm unpol}-\frac{1}{2}Re\Bigg[
  \left(\eta{\rm e}^{-i\phi}+\bar{\eta}{\rm e}^{i\phi)}\right)
  {\rm e}^{-i\chi}\Sigma_{02}\Bigg]\nonumber\\
&&+\frac{1}{2}Re\Bigg[
  \left(\eta{\rm e}^{-i\phi}-\bar{\eta}{\rm e}^{i\phi}\right)
  {\rm e}^{-i\chi}\Delta_{02}\Bigg]
  +\eta\bar{\eta}Re\Bigg[
  {\rm e}^{-2i\phi}\Sigma_{22}+{\rm e}^{-2i\chi}\Sigma_{00}
  \Bigg],\\
&=&\Sigma_{\rm unpol}
  -\frac{1}{2}[\eta\cos(\phi+\chi)+\bar{\eta}\cos(\phi-\chi)]
   {\cal R}(\Sigma_{02})\nonumber\\  
&&\hskip 1.5cm 
  +\frac{1}{2}[\eta\sin(\phi+\chi)-\bar{\eta}\sin(\phi-\chi)]
   {\cal I}(\Sigma_{02})\nonumber\\  
&&\hskip 1.5cm 
  -\frac{1}{2}[\eta\cos(\phi+\chi)-\bar{\eta}\cos(\phi-\chi)]
   {\cal R}(\Delta_{02})\nonumber\\  
&&\hskip 1.5cm 
  +\frac{1}{2}[\eta\sin(\phi+\chi)+\bar{\eta}\sin(\phi-\chi)]
   {\cal I}(\Delta_{02})\nonumber\\  
&&\hskip 1.5cm
  +\eta\bar{\eta}\cos(2\phi){\cal R}(\Sigma_{22})
  +\eta\bar{\eta}\sin(2\phi){\cal I}(\Sigma_{22})\nonumber\\
&&\hskip 1.5cm
  +\eta\bar{\eta}\cos(2\chi){\cal R}(\Sigma_{00})
  +\eta\bar{\eta}\sin(2\chi){\cal I}(\Sigma_{00}),
 \label{linear dist}
\end{eqnarray}
where the invariant functions are defined as
\setcounter{equation}{12}
\renewcommand{\theequation}{\arabic{section}.\arabic{equation}}
\begin{eqnarray}
&&\Sigma_{\rm unpol}=\frac{1}{4}\sum_X
              \left[|M_{++}|^2+|M_{+-}|^2
                   +|M_{-+}|^2+|M_{--}|^2\right]\nonumber\\
&&\Sigma_{02}=\frac{1}{2}\sum_X\left[M_{++}(M^*_{+-}+M^*_{-+})
             +(M_{+-}+M_{-+})M^*_{--}\right]\nonumber\\
&&\Delta_{02}=\frac{1}{2}\sum_X\left[M_{++}(M^*_{+-}-M^*_{-+})
             -(M_{+-}-M_{-+})M^*_{--}\right]\nonumber\\
&&\Sigma_{22}=\frac{1}{2}\sum_X(M_{+-}M^*_{-+}),\qquad
  \Sigma_{00}=\frac{1}{2}\sum_X(M_{++}M^*_{--}),
\end{eqnarray}
with the subscripts, $0$ and $2$, representing the magnitude of the 
sum of the helicities of the initial two-photon system.

\subsection{Symmetry properties}

It is useful to classify the invariant functions according to 
their transformation properties under the discrete symmetries,
CP and CP$\tilde{\rm T}$\cite{KhRp}.
We find that CP invariance leads to the relations
\setcounter{equation}{0}
\renewcommand{\theequation}{\arabic{section}.14\alph{equation}}
\begin{eqnarray}
\sum_X\left(M_{\lambda_1\lambda_2}
            M^*_{\lambda^\prime_1\lambda^\prime_2}\right)
&=&
\sum_X\left(M_{-\lambda_2,-\lambda_1}
            M^*_{-\lambda^\prime_2,-\lambda^\prime_1}\right),\\
{\rm d}\sigma(\phi,\chi;\eta,\bar{\eta})&=&
{\rm d}\sigma(\phi,-\chi;\bar{\eta},\eta),
\end{eqnarray}
and, if there are no absorptive parts in the amplitudes, 
CP$\tilde{\rm T}$ invariance leads to the realtaions
\setcounter{equation}{0}
\renewcommand{\theequation}{\arabic{section}.15\alph{equation}}
\begin{eqnarray}
\sum_X\left(M_{\lambda_1\lambda_2}
            M^*_{\lambda^\prime_1\lambda^\prime_2}\right)
&=&
\sum_X\left(M^*_{-\lambda_2,-\lambda_1}
            M_{-\lambda^\prime_2,-\lambda^\prime_1}\right),\\
{\rm d}\sigma(\phi,\chi;\eta,\bar{\eta})&=&
{\rm d}\sigma(-\phi,\chi;\bar{\eta},\eta).
\end{eqnarray}

The nine invariant functions in Eq.~(\ref{linear dist}) 
can then be divided into four categories
under CP and CP$\tilde{\rm T}$: even-even, even-odd, odd-even,
and odd-odd terms as in Table~1. CP-odd coefficients directly measure 
CP violation and CP$\tilde{\rm T}$-odd terms indicate rescattering 
effects (absorptive parts in the scattering amplitudes). Table~1 shows 
that there exist three CP-odd functions; ${\cal I}(\Sigma_{02})$, 
${\cal I}(\Sigma_{00})$ and ${\cal R}(\Delta_{02})$. Here, ${\cal R}$
and ${\cal I}$ are for real and imaginary parts, respectively.
While the first two terms are CP$\tilde{\rm T}$-even, the last 
term ${\cal R}(\Delta_{02})$ is
CP$\tilde{\rm T}$-odd. Since the CP$\tilde{\rm T}$-odd term 
${\cal R}(\Delta_{02})$ requires the absorptive part in the amplitude,
it is generally expected to be smaller in magnitude than 
the CP$\tilde{\rm T}$-even terms. We therefore study the two CP-odd and 
CP$\tilde{\rm T}$-even distributions;
${\cal I}(\Sigma_{02})$ and ${\cal I}(\Sigma_{00})$.
\setcounter{equation}{15}
\renewcommand{\theequation}{\arabic{section}.\arabic{equation}}
\begin{enumerate}
\item[{ }] TABLE I. CP and CP$\tilde{\rm T}$ properties 
      of the invariant functions and the angular distributions.  
\end{enumerate}
\vskip 0.1cm
\begin{center}
\begin{tabular}{|c|c|c|c|}\hline
 \mbox{ }\hskip 0.2cm CP\mbox{ }\hskip 0.2cm   
&\mbox{ }\hskip 0.2cm CP$\tilde{\rm T}$\mbox{ }\hskip 0.2cm   
&\mbox{ }\hskip 0.2cm Invariant functions \mbox{ }\hskip 0.2cm  
&\mbox{ }\hskip 0.2cm Angular dependences \mbox{ }\hskip 0.2cm  \\
\hline
even & even & $\Sigma_{\rm unpol}$     
            & { } \\
            \cline{3-4}
{ }  & { }  & ${\cal R}(\Sigma_{02})$  
            & $\eta\cos(\phi+\chi)+\bar{\eta}\cos(\phi-\chi)$  \\
            \cline{3-4}
{ }  & { }  & ${\cal R}(\Sigma_{22})$
            & $\eta\bar{\eta}\cos(2\phi)$ \\
            \cline{3-4}
{ }  & { }  & ${\cal R}(\Sigma_{00})$  
            & $\eta\bar{\eta}\cos(2\chi)$ \\ \hline 
even & odd  & ${\cal I}(\Delta_{02})$
            & $\eta\sin(\phi+\chi)+\bar{\eta}\sin(\phi-\chi)$  \\
            \cline{3-4}
{ }  & { }  & ${\cal I}(\Sigma_{22})$  
            & $\eta\bar{\eta}\sin(2\phi)$ \\ \hline
odd  & even & ${\cal I}(\Sigma_{02})$
            & $\eta\sin(\phi+\chi)-\bar{\eta}\sin(\phi-\chi)$  \\
            \cline{3-4}
{ }  & { }  & ${\cal I}(\Sigma_{00})$  
            & $\eta\bar{\eta}\sin(2\chi)$  \\ \hline
odd  & odd  & ${\cal R}(\Delta_{02})$  
            & $\eta\cos(\phi+\chi)-\bar{\eta}\cos(\phi-\chi)$  \\
\hline
\end{tabular}
\end{center}

\vskip 0.3cm
We can define two CP-odd asymmetries from the two distributions, 
${\cal I}(\Sigma_{02})$ and ${\cal I}(\Sigma_{00})$. 
First, we note that the $\Sigma_{00}$ term does not depend on the 
azimuthal angle $\phi$ whereas the $\Sigma_{02}$ does. 
In order to improve the observability we may integrate 
the ${\cal I}(\Sigma_{02})$ term over the azimuthal angle $\phi$ 
with an appropriate weight function. 
Without any loss of generality we can take $\eta=\bar{\eta}$. 
Then, the quantity ${\cal I}(\Sigma_{00})$ in Eq.~(\ref{linear dist}) 
can be separated by taking the difference of the distributions 
at $\chi=\pm\pi/4$  and the ${\cal I}(\Sigma_{02})$ by taking 
the difference of the distributions at $\chi=\pm\pi/2$.
As a result we obtain the following two integrated CP-odd
asymmetries:
\begin{eqnarray}
\hat{A}_{02}=\left(\frac{2}{\pi}\right)
     \frac{{\cal I}(\Sigma_{02})}{\Sigma_{\rm unpol}},\qquad
\hat{A}_{00}=\frac{{\cal I}(\Sigma_{00})}{\Sigma_{\rm unpol}},
\end{eqnarray}
where the factor $(2/\pi)$ in the $\hat{A}_{02}$ stems from taking 
the average over the azimuthal angle $\phi$ with the weight function
${\rm sign}(\cos\phi)$:
\setcounter{equation}{0}
\renewcommand{\theequation}{\arabic{section}.17\alph{equation}}
\begin{eqnarray}
&&\hat{A}_{02}=\frac{\int^{4\pi}_0{\rm d}\phi [{\rm sign}(\cos\phi)]
\bigg[\left(\frac{{\rm d}\sigma}{{\rm d}\phi}\right)_{\chi=\frac{\pi}{2}}
    -\left(\frac{{\rm d}\sigma}{{\rm d}\phi}\right)_{\chi=-\frac{\pi}{2}}
\bigg]}{\int^{4\pi}_0{\rm d}\phi 
\bigg[\left(\frac{{\rm d}\sigma}{{\rm d}\phi}\right)_{\chi=\frac{\pi}{2}}
    +\left(\frac{{\rm d}\sigma}{{\rm d}\phi}\right)_{\chi=-\frac{\pi}{2}}
  \bigg]},\\
&&\hat{A}_{00}=\frac{\int^{4\pi}_0{\rm d}\phi
\bigg[\left(\frac{{\rm d}\sigma}{{\rm d}\phi}\right)_{\chi=\frac{\pi}{4}}
    -\left(\frac{{\rm d}\sigma}{{\rm d}\phi}\right)_{\chi=-\frac{\pi}{4}}
\bigg]}{\int^{4\pi}_0{\rm d}\phi 
\bigg[\left(\frac{{\rm d}\sigma}{{\rm d}\phi}\right)_{\chi=\frac{\pi}{4}}
    +\left(\frac{{\rm d}\sigma}{{\rm d}\phi}\right)_{\chi=-\frac{\pi}{4}}
\bigg]}.
\end{eqnarray}
In pair production processes such as 
$\gamma\gamma\rightarrow W^+W^-$, all the distributions, $\Sigma_i$,
can be integrated over the scattering angle $\theta$ with a CP-even 
angular cut so as to test CP violation.
  
\setcounter{equation}{0}
\renewcommand{\theequation}{\arabic{section}.\arabic{equation}}

\vskip 1.5cm

\section{Photon linear collider}
\label{sec:PLC}

In this section we give a short review of the powerful 
mechanism of providing an energetic, highly polarized photon beam;
the Compton laser backscattering\cite{GKS} off energetic electron or
positron beams. After the review we introduce two functions 
to describe partial linear polarization transfer from 
the laser beams to the backscattered photon beams for the 
photon-photon collisions.

We assume that the electron or 
positron beams are unpolarized and the laser beams are purely
linearly polarized. Even in that case the backscattered photon 
beam is not purely linearly polarized but only part of the laser 
linear polarization is transferred to the backscattered energetic
high-energy photon beam. 

\subsection{Photon spectrum}

We consider the situation where a purely linearly polarized laser beam 
of frequency $\omega_0$ is focused upon an unpolarized electron or 
positron beam of energy $E$. 
In the collision of a laser photon beam and a linac electron beam, 
a high energy photon beam of energy $\omega$, which is partially 
linearly polarized, is emitted at a very small angle, along with the 
scattered electron beam of energy $E^\prime =E-\omega$. 
The kinematics of the Compton backscattering process is then 
characterized by the dimensionless parameters $x$ and $y$:
\begin{eqnarray}
x=\frac{4E\omega_0}{m^2_e}
 \approx 15.3\left(\frac{E}{\rm TeV}\right)
             \left(\frac{\omega_0}{\rm eV}\right),\qquad
y=\frac{\omega}{E}.
\end{eqnarray}
In general, the backscattered photon energies increase with $x$;
the maximum photon energy fraction is given by
$y_m=x/(1+x)$. Operation below the threshold\cite{GKS} for $e^+e^-$ 
pair production in collisions between the laser beam and the 
Compton-backscattered photon beam requires $x\leq 2(1+\sqrt{2})\approx
4.83$; the lower bound on $x$ depends on the lowest available
laser frequency and the production threshold of a given final
state. 

The backscattered photon energy spectrum is given by the function
\begin{eqnarray}
\phi_0(y)=\frac{1}{1-y}+1-y-4r(1-r),
\end{eqnarray}
where $r=y/(x(1-y))$. Fig.2(a) shows the photon energy spectrum
for various values of $x$. Clearly large values of $x$ are
favored to produce highly energetic photons.
On the other hand, the degree of linear polarization of the 
backscattered photon beam is given by\cite{GKS} 
\begin{eqnarray}
\eta(y)=\frac{2r^2}{\phi_0(y)}.
\end{eqnarray}
The maximum linear polarization is reached for $y=y_m$ (See Fig.~2(b)),
\begin{eqnarray}
\eta_{\rm max}=\eta(y_m)=\frac{2(1+x)}{1+(1+x)^2},
\end{eqnarray}
and approaches unity for small values of $x$. 
In order to retain large linear polarization  we should keep
the $x$ value as small as possible.

\subsection{Linear polarization transfers}

In the two-photon collision case only part of each
laser linear polarization is transferred to the high-energy photon beam.
We introduce  two functions, ${\cal A}_\eta$ and ${\cal A}_{\eta\eta}$, 
to denote the degrees of linear polarization transfer\cite{Comment2} 
as
\begin{eqnarray}
{\cal A}_\eta(\tau)=\frac{\langle \phi_0\phi_3\rangle_\tau}{\langle 
               \phi_0\phi_0\rangle_\tau},\qquad 
{\cal A}_{\eta\eta}(\tau)
     =\frac{\langle \phi_3\phi_3\rangle_\tau}{\langle 
               \phi_0\phi_0\rangle_\tau},
\end{eqnarray}
where $\phi_3(y)=2r^2$ and $\tau$ is the ratio of the $\gamma\gamma$ 
c.m. energy squared $\hat{s}$ to the $e^+e^-$ collider energy squared 
$s$. The function ${\cal A}_\eta$ is for the collision of an 
unpolarized photon beam and a linearly polarized photon beam and 
the function ${\cal A}_{\eta\eta}$ for the collision of two linearly 
polarized photon beams. The convolution integrals 
$\langle \phi_i\phi_j\rangle_\tau$ ($i,j=0,3$) 
for a fixed value of $\tau$ are defined as  
\begin{eqnarray}
\langle \phi_i\phi_j\rangle_\tau
 =
 \frac{1}{{\cal N}^2(x)}\int^{y_m}_{\tau/y_m}\frac{{\rm d}y}{y}
 \phi_i(y)\phi_j(\tau/y),
\end{eqnarray}
where the normalization factor ${\cal N}(x)$ is given by the integral 
of the photon energy spectrum $\phi_0$ over $y$ as
\begin{eqnarray}
{\cal N}(x)=\int^{y_m}_0\phi_0(y){\rm d}y
    =\ln(1+x)\left[1-\frac{4}{x}-\frac{8}{x^2}\right]
      +\frac{1}{2}+\frac{8}{x}-\frac{1}{2(1+x)^2}.
\end{eqnarray}

The event rates of the $\gamma\gamma\rightarrow X$ reaction with 
polarized photons can be obtained by folding a photon luminosity 
spectral function with the $\gamma\gamma\rightarrow X$ production 
cross section as (for $\eta=\bar{\eta}$) 
\begin{eqnarray}
{\rm d}N_{\gamma\gamma\rightarrow X}
 ={\rm d}L_{\gamma\gamma}\cdot 
  {\rm d}\hat{\sigma}(\gamma\gamma\rightarrow X),
\end{eqnarray}
where
\setcounter{equation}{0}
\renewcommand{\theequation}{\arabic{section}.9\alph{equation}}
\begin{eqnarray}
{\rm d}L_{\gamma\gamma}
  &=&\kappa^2L_{ee}\langle\phi_0\phi_0\rangle_\tau{\rm d}\tau,\\
{\rm d}\hat{\sigma}(\gamma\gamma\rightarrow X)&=&  
   \frac{1}{2\hat{s}}{\rm d}\Phi_X
  \Bigg[\Sigma_{\rm unpol}-\eta{\cal A}_\eta\cos\phi 
        Re\bigg({\rm e}^{-i\chi}\Sigma_{02}\bigg)\nonumber\\
 &&
 +\eta{\cal A}_\eta\sin\phi Im\bigg({\rm e}^{-i\chi}\Delta_{02}\bigg)
 +\eta^2{\cal A}_{\eta\eta} 
  Re\bigg({\rm e}^{-2i\phi}\Sigma_{22}
         +{\rm e}^{-2i\chi}\Sigma_{00}\bigg)\Bigg].
\label{folded dist}
\end{eqnarray}
Here, $\kappa$ is the $e$-$\gamma$ conversion coefficient in
the Compton backscattering and  ${\rm d}\Phi_X$ is the phase space
factor of the final state, which is given for $X=W^+W^-$ by
\setcounter{equation}{9}
\renewcommand{\theequation}{\arabic{section}.\arabic{equation}}
\begin{eqnarray}
{\rm d}\Phi_{W^+W^-}=\frac{\hat{\beta}}{32\pi^2}
                    {\rm d}\cos\hat{\theta}{\rm d}\phi.
\end{eqnarray}
The distribution (\ref{folded dist}) of event rates enables us to 
construct two CP-odd asymmetries;
\begin{eqnarray}
A_{02}=\left(\frac{2}{\pi}\right)\frac{N_{02}}{N_{\rm unpol}},\qquad
A_{00}=\frac{N_{00}}{N_{\rm unpol}},
\end{eqnarray}
where with $\tau_{\rm max}=y_m^2$ and $\tau_{\rm min}=M^2_X/s$
we have for the event distributions 
\begin{eqnarray}
  \left(\begin{array}{c}
      N_{\rm upl} \\
      N_{02}\\
      N_{00}\end{array}\right)
    =\kappa^2L_{ee}\frac{1}{2s}
     \int^{\tau_{\rm max}}_{\tau_{\rm min}}\frac{{\rm d}\tau}{\tau}
     \int {\rm d}\Phi_X 
     \langle\phi_0\phi_0\rangle_\tau
   \left(\begin{array}{c}
     \Sigma_{\rm unpol}\\
     \eta{\cal A}_\eta{\cal I}\left(\Sigma_{02}\right)\\
     \eta^2{\cal A}_{\eta\eta}{\cal I}\left(\Sigma_{00}\right)
       \end{array}\right).
\end{eqnarray}
The asymmetries depend crucially on the two-photon spectrum
and the two linear polarization transfers. 

We first investigate the $\sqrt{\tau}$ dependence of the two-photon 
spectrum and the two linear polarization transfers,
$A_\eta$ and $A_{\eta\eta}$ by varying the value of the dimensionless 
parameter $x$. 
Three values of $x$ are chosen; $x=0.5$, $1$, and $4.83$.
Two figures in Fig.~3 clearly show that the energy of two photons 
reaches higher ends for larger $x$ values but the maximum linear 
polarization transfers are larger for smaller $x$ values.
We also note that $A_\eta$ (solid lines) is larger than $A_{\eta\eta}$ 
(dashed lines) in the whole range of $\sqrt{\tau}$. 
We should keep the parameter $x$ as large as possible to
reach higher energies. However, larger CP-odd asymmetries can be
obtained for smaller $x$ values. Therefore, there should
exist a compromised value of $x$ for the optimal observality of 
CP violation. The energy dependence of the subprocess cross section
and that of the CP-odd asymmetries are both essential to find
the optimal $x$ value.
\vskip 1.5cm

\section{CP-odd weak-boson couplings}
\label{sec:model}

In this section we describe how CP violation from new interactions
among electroweak vector bosons can be probed
in a model-independent way in the $W$ pair production in two-photon
collisions. 
We adopt the effective Lagrangian with most general CP-odd interactions
among electroweak gauge bosons. The basic assumptions are that the 
operators with lowest energy dimension (6) dominate the CP-odd 
amplitudes and that they respect the electroweak gauge invariance
which is broken spontaneously by an effective SU(2)-doublet scalar.
The effective Lagrangian then determines the energy dependence of the
scattering amplitudes at energies below the new physics scale.

\subsection{Effective Lagrangian with CP-odd operators}

The effects of new physics are parametrized by using an effective 
Lagrangian in a model and process independent way. 
As for the electroweak gauge symmetry breaking parameter, we adopt the
effective SU(2)-doublet scalar field $\Phi$, which is more convenient
when a physical Higgs boson appears at low energies.
In addition to the Higgs doublet field $\Phi$, the building blocks of 
the gauge-invariant operators are the covariant derivatives of the Higgs
field, $D_\mu\Phi$, and the non-Abelian-field strength tensors
$W^I_{\mu\nu}$ ($I=1,2,3$) and $B_{\mu\nu}$ of the SU(2)$_{\rm L}$
and U(1)$_{\rm Y}$ gauge fields, respectively.
Considering CP-odd interactions of dimension six, we can construct
six independent operators that are relevant for the process
$\gamma\gamma\rightarrow W^+W^-$ 
\renewcommand{\theequation}{\arabic{section}.1\alph{equation}}
\begin{eqnarray}
&&{\cal O}_{B\tilde{B}}=g^{\prime 2}(\Phi^\dagger\Phi)
                        B_{\mu\nu}\tilde{B}^{\mu\nu},\\
&&{\cal O}_{B\tilde{W}}=gg^\prime(\Phi^\dagger\sigma^I\Phi)
                        B_{\mu\nu}\tilde{W}^{I\mu\nu},\\
&&{\cal O}_{W\tilde{W}}
   =g^2(\Phi^\dagger\Phi)W^I_{\mu\nu}\tilde{W}^{I\mu\nu},\\
&&{\cal O}_{\tilde{B}}
   =ig^\prime\left[(D_\mu\Phi)^\dagger(D_\nu\Phi)\right]
       \tilde{B}^{\mu\nu},\\
&&{\cal O}_{\tilde{W}}
   =ig\left[(D_\mu\Phi)^\dagger\sigma^I(D_\nu\Phi)\right]
      \tilde{W}^{I\mu\nu},\\
&&{\cal O}_{WW\tilde{W}}=g^3\epsilon^{IJK}\tilde{W}^{I\mu\nu}
                       W_\nu^{J\rho}W^K_{\rho\mu},
\label{new operators}
\end{eqnarray}
where $\tilde{W}^{I\mu\nu}=\frac{1}{2}\epsilon^{\mu\nu\alpha\beta}
W^I_{\alpha\beta}$, $\tilde{B}^{\mu\nu}
=\frac{1}{2}\epsilon^{\mu\nu\alpha\beta}
\tilde{B}_{\alpha\beta}$, $\sigma^I$ are the Pauli matrices, and 
the SU(2)$_{\rm L}\times$U(1)$_{\rm Y}$ covariant derivative 
is given by
\setcounter{equation}{1}
\renewcommand{\theequation}{\arabic{section}.\arabic{equation}}
\begin{eqnarray}
D_\mu=\partial_\mu+ig\frac{\sigma^I}{2}W^I_\mu 
                  +ig^\prime Y B_\mu,
\end{eqnarray}
with the isospin indices $I,J$ and $K$ ($=1,2,3$) 
and the SU(2) and U(1) couplings, $g$ and $g^\prime$, respectively.
The effective Lagrangian is written as 
\begin{eqnarray}
{\cal L}={\cal L}_{SM}&+&\frac{1}{\Lambda^2}
           \bigg[f_{B\tilde{B}}{\cal O}_{B\tilde{B}}
               + f_{B\tilde{W}}{\cal O}_{B\tilde{W}}
               + f_{W\tilde{W}}{\cal O}_{W\tilde{W}}\nonumber\\
             &+& f_{\tilde{B}}{\cal O}_{\tilde{B}}
               + f_{\tilde{W}}{\cal O}_{\tilde{W}}
               + f_{WW\tilde{W}}{\cal O}_{WW\tilde{W}}\bigg],
\label{effective}
\end{eqnarray}
where the dimension-six terms ${\cal O}_i$ are scaled by the common 
dimensional parameter $\Lambda$ with dimensionless coefficients $f_i$.
The fields $W^3$ and $B$ are related in terms of the 
Weinberg angle $\theta_W$ to the $Z$ and photon fields, $Z$ and $A$
as
\begin{eqnarray}
\left(\begin{array}{c}
      W^3\\
      B\end{array}\right)
 =\left(\begin{array}{cc}
        \cos\theta_W & \sin\theta_W \\
       -\sin\theta_W & \cos\theta_W
        \end{array}\right)
   \left(\begin{array}{c}
         Z \\
         A\end{array}\right).
\end{eqnarray}
Incidentally, as we are interested in the photon-induced process 
$\gamma\gamma\rightarrow W^+W^-$, we can neglect the terms
involving the $Z$ field. Then all the terms for the process
$\gamma\gamma\rightarrow W^+W^-$ can be derived by the following 
effective replacements 
\setcounter{equation}{0}
\renewcommand{\theequation}{\arabic{section}.5\alph{equation}}
\begin{eqnarray}
W^-_{\mu\nu}&\rightarrow &(\partial_\mu-ieA_\mu)W^-_\nu
             -(\partial_\nu-ieA_\nu)W^-_\mu,\\
W^+_{\mu\nu}&\rightarrow &(\partial_\mu+ieA_\mu)W^+_\nu
             -(\partial_\nu+ieA_\nu)W^+_\mu,\\
W^3_{\mu\nu}&\rightarrow &\sin\theta_WF_{\mu\nu}
             +\frac{ie}{\sin\theta_W}
              (W^+_\mu W^-_\nu-W^+_\nu W^-_\mu),\\
B_{\mu\nu}&\rightarrow &\cos\theta_W F_{\mu\nu},
\end{eqnarray}
where $F_{\mu\nu}=\partial_\mu A_\nu-\partial_\nu A_\mu$.
We take the unitary gauge where the scalar doublet $\Phi$ 
with hypercharge $Y=\frac{1}{2}$ takes the form
\setcounter{equation}{5}
\renewcommand{\theequation}{\arabic{section}.\arabic{equation}}
\begin{eqnarray}
\Phi=\frac{1}{\sqrt{2}}(v+H)\left[\begin{array}{c}
                           0\\
                           1      \end{array}\right].
\label{unitary}
\end{eqnarray}
$H$ denotes the Higgs boson in the SM.
It is now straightforward to obtain the new CP-odd vertices 
among terms of the component fields, $W^\pm$, $A$, and $H$ in
the unitary gauge.

\subsection{CP-odd vertices}

In this section we give  the Feynman rules for
the $\gamma WW$, $\gamma\gamma H$, $HWW$, and $\gamma\gamma WW$
vertices, relevant for the $\gamma\gamma\rightarrow W^+W^-$ reaction.
Table~2 shows which vertices already exist in the SM at tree level 
and which  new vertices appear from the new dimension-six CP-odd 
operators. Firstly, the three operators, ${\cal O}_{B\tilde{B}}$, 
${\cal O}_{B\tilde{W}}$, and ${\cal O}_{W\tilde{W}}$ 
contribute to the $\gamma\gamma H$ vertex.
Secondly, we find that the operator
${\cal O}_{WW\tilde{W}}$ gives a new CP-odd $\gamma WW$ vertex and
a new CP-odd $\gamma\gamma WW$ vertex, which are related by U(1) 
electromagnetic gauge invariance.
In addition, the three operators, ${\cal O}_{B\tilde{W}}$,
${\cal O}_{\tilde{W}}$, and ${\cal O}_{\tilde{B}}$ contribute to the
$\gamma WW$ vertex as well. Thirdly, we find that ${\cal O}_{W\tilde{W}}$
and ${\cal O}_{\tilde{W}}$ contribute to the $HWW$ vertex.
\begin{enumerate}
\item[{ }] TABLE II. Vertices relevant for the process
          $\gamma\gamma\rightarrow W^+W^-$ in the SM with 
          dimension-six CP-odd terms.    
\end{enumerate}
\begin{center}
\begin{tabular}{|c|c|c|c|c|}\hline
 \mbox{ }\hskip 0.2cm Vertex\mbox{ }\hskip 0.2cm 
&\mbox{ }\hskip 0.2cm $\gamma WW$ \mbox{ }\hskip 0.2cm 
&\mbox{ }\hskip 0.2cm $\gamma\gamma WW$ \mbox{ }\hskip 0.2cm 
&\mbox{ }\hskip 0.2cm $HWW$ \mbox{ }\hskip 0.2cm 
&\mbox{ }\hskip 0.2cm $\gamma\gamma H$ \mbox{ }\hskip 0.2cm\\ 
\hline
SM &  $\bigcirc$  &  $\bigcirc$  &  $\bigcirc$  &  X  \\ 
\hline
\mbox{ }\hskip 0.2cm ${\cal O}_{B\tilde{B}}$\mbox{ }\hskip 0.2cm 
   &  X  &  X  &  X  &  $\bigcirc$       \\ 
\hline
\mbox{ }\hskip 0.2cm ${\cal O}_{B\tilde{W}}$\mbox{ }\hskip 0.2cm 
   &  $\bigcirc$  &  X  &  X  &  $\bigcirc$       \\ 
\hline
\mbox{ }\hskip 0.2cm ${\cal O}_{W\tilde{W}}$\mbox{ }\hskip 0.2cm
   &  X  &  X  &  $\bigcirc$  &  $\bigcirc$       \\ 
\hline
\mbox{ }\hskip 0.2cm ${\cal O}_{\tilde{B}}$\mbox{ }\hskip 0.2cm 
   &  $\bigcirc$  &  X  &  X  &  X   \\ 
\hline
\mbox{ }\hskip 0.2cm ${\cal O}_{\tilde{W}}$\mbox{ }\hskip 0.2cm 
   &  $\bigcirc$  &  X  &  $\bigcirc$  & X      \\ 
\hline
\mbox{ }\hskip 0.2cm ${\cal O}_{WW\tilde{W}}$\mbox{ }\hskip 0.2cm 
   &  $\bigcirc$  & $\bigcirc$  & X &  X       \\ 
\hline
\end{tabular}
\end{center}
\vskip 0.3cm

For convenience we define four new dimensionless form factors, 
$Y_i$ ($i=1$ to $4$), which are related with the coefficients, $f_i$'s 
($i=B\tilde{B},B\tilde{W},W\tilde{W},\tilde{B},\tilde{W},WW\tilde{W}$) 
as
\begin{eqnarray}
Y_1&=&\left(\frac{m_W}{\Lambda}\right)^2
      \bigg[f_{B\tilde{W}}
         +\frac{1}{4}f_{\tilde{B}}+f_{\tilde{W}}\bigg],\qquad 
Y_2 = \left(\frac{m_W}{\Lambda}\right)^2
      \frac{g^2}{4}f_{WW\tilde{W}},\nonumber\\
Y_3&=&\left(\frac{m_W}{\Lambda}\right)^2
      \bigg[f_{W\tilde{W}}+\frac{1}{4}f_{\tilde{W}}\bigg],\qquad
Y_4 = \left(\frac{m_W}{\Lambda}\right)^2
      \bigg[f_{B\tilde{B}}-f_{B\tilde{W}}-f_{W\tilde{W}}\bigg].
\end{eqnarray}
If all the coefficients, $f_i$, are of the similar size, 
then $Y_2$ would be about ten times smaller than the other form 
factors in size because of the factor $g^2/4\sim 0.1$.
We denote the Feynman rule of a vertex ${\rm V}$ in the form;
$ie\Gamma_{\rm V}$.
It is then straightforward to derive the explicit form of two 
simple $\gamma\gamma H$ and $HWW$ vertices; 
\begin{eqnarray}
\Gamma^{\mu\nu}_{\gamma\gamma H}(k_1,k_2)
 &=&
\frac{8Y_4}{m_W}\sin\theta_W\epsilon^{\mu\nu\rho\sigma}
 k_{1\rho}k_{2\sigma},\\
\Gamma^{\alpha\beta}_{HWW}(q_1,q_2)
 &=& 
\frac{m_W}{\sin\theta_W}g^{\alpha\beta}
+\frac{8Y_3}{m_W\sin\theta_W}\epsilon^{\alpha\beta\rho\sigma}
  q_{1\rho}q_{2\sigma}, 
\end{eqnarray}
where $k_1(\mu)$ and $k_2(\nu)$ are four-momenta (Lorentz indices) of 
two incoming photons and $q_1(\alpha)$ and $q_2(\beta)$ are four-momenta 
(Lorentz indices) for the outgoing $W^+$ and $W^-$, respectively. 
In the SM the $\gamma\gamma H$ vertex appears in the one-loop level,
we do not study its consequences in this paper.
The triple $\gamma WW$ vertex is 
\begin{eqnarray}
\Gamma^{\mu\alpha\beta}_{\gamma WW}(k,q_1,q_2)
  &=&
(q_1-q_2)^\mu g^{\alpha\beta}
-(q_1+k)^\beta g^{\mu\alpha}
+(k+q_2)^\alpha g^{\mu\beta}\nonumber\\
&-&\!4Y_1\epsilon^{\mu\alpha\beta\rho}k_\rho\nonumber\\
&+&\! 12\frac{Y_2}{m^2_W}\bigg[
   2(q_1\cdot q_2)\epsilon^{\mu\alpha\beta\rho}
  \! -q^\alpha_2\epsilon^{\mu\beta\rho\sigma}(q_1-q_2)_\sigma
  \! -q^\beta_1\epsilon^{\mu\alpha\rho\sigma}(q_1-q_2)_\sigma
  \bigg]k_\rho,
\end{eqnarray}
where $k=q_1+q_2$, and the quartic $\gamma\gamma WW$ vertex is   
\begin{eqnarray}
&& \Gamma^{\mu\nu\alpha\beta}_{\gamma\gamma WW}(k_1,k_2,q_1,q_2)
 =
-2g^{\mu\nu}g^{\alpha\beta}
 +g^{\mu\alpha}g^{\nu\beta} 
 +g^{\mu\beta}g^{\nu\alpha}\nonumber\\
&& \mbox{ }\hskip 1.2cm +8\frac{Y_2}{m^2_W}\Bigg[
    2g^{\alpha\beta}\epsilon^{\mu\nu\rho\sigma}k_{1\rho}k_{2\sigma}
   +2g^{\mu\nu}\epsilon^{\alpha\beta\rho\sigma}
     q_{1\rho}q_{2\sigma}\nonumber\\
 && { }\hskip 2.4cm
   -g^{\mu\alpha}\epsilon^{\nu\beta\rho\sigma}q_{2\rho}k_{2\sigma}
   -g^{\mu\beta}\epsilon^{\nu\alpha\rho\sigma}
    q_{1\rho}k_{2\sigma}\nonumber\\
 && { }\hskip 2.4cm
   -g^{\nu\alpha}\epsilon^{\mu\beta\rho\sigma}q_{2\rho}k_{1\sigma}
   -g^{\nu\beta}\epsilon^{\mu\alpha\rho\sigma}q_{1\rho}k_{1\sigma}
   +(k_1-k_2)\cdot (q_1-q_2)\epsilon^{\mu\nu\alpha\beta}\nonumber\\
 && { }\hskip 2.4cm
   +k^\mu_2\epsilon^{\nu\alpha\beta\rho}(q_1-q_2)_\rho
   +k^\nu_1\epsilon^{\mu\alpha\beta\rho}(q_1-q_2)_\rho\nonumber\\
 && { }\hskip 2.4cm
   +q^\alpha_2\epsilon^{\mu\nu\beta\rho}(k_1-k_2)_\rho
   +q^\beta_1\epsilon^{\mu\nu\alpha\rho}(k_1-k_2)_\rho\nonumber\\
 && { }\hskip 2.4cm
   +(q_1-q_2)^\mu\epsilon^{\nu\alpha\beta\rho}(k_1+k_2)_\rho
   +(q_1-q_2)^\nu\epsilon^{\mu\alpha\beta\rho}(k_1+k_2)_\rho\nonumber\\
 && { }\hskip 2.4cm
   +(k_1-k_2)^\alpha\epsilon^{\mu\nu\beta\rho}(q_1+q_2)_\rho
   +(k_1-k_2)^\beta\epsilon^{\mu\nu\alpha\rho}(q_1+q_2)_\rho\Bigg].
\end{eqnarray}
\vskip 1.5cm

\section{Helicity amplitudes for $\gamma\gamma\rightarrow W^+W^-$}
\label{sec:amplitude}

In this section we present the complete calculation of
polarization amplitudes for the process
\begin{eqnarray}
\gamma(k_1,\lambda_1)+\gamma(k_2,\lambda_2)
 \rightarrow 
W^+(q_1,\lambda_3)+W^-(q_2,\lambda_4),
\end{eqnarray}
with the effective Lagrangian (\ref{effective}) in Section~IV.
The four-momentum and the helicity of each particle are 
shown in the parenthesis. The helicities of the $W$ are given
in the $\gamma\gamma$ c.m. frame.
Helicity amplitudes contain full information of the process. 
The relative phases of the amplitudes are essential
because the interference of different photon and $W$ helicity
states gives a nontrivial azimuthal-angle dependence.

By taking the two photon momenta along the $z$-axis and by taking the
$W^+$ momentum in the $x$-$z$ plane (see Fig.~1),
the four-momenta are parametrized as
\begin{eqnarray}
&&k^\mu_1=\frac{\sqrt{\hat{s}}}{2}(1,0,0,1),\qquad
  k^\mu_2=\frac{\sqrt{\hat{s}}}{2}(1,0,0,-1),\nonumber\\
&&q^\mu_1=\frac{\sqrt{\hat{s}}}{2}
    \left(1,\hat{\beta}\sin\theta,0,\hat{\beta}\cos\theta\right),\qquad
  q^\mu_2=\frac{\sqrt{\hat{s}}}{2}
    \left(1,-\hat{\beta}\sin\theta,0,-\hat{\beta}\cos\theta\right).
\end{eqnarray}
The incoming photon polarization vectors are
\begin{eqnarray}
\epsilon^\mu_1(\pm)=\mp\frac{1}{\sqrt{2}}(0,1,\pm i,0),\qquad
\epsilon^\mu_2(\pm)=\mp\frac{1}{\sqrt{2}}(0,1,\mp i,0),
\end{eqnarray}
and the transverse (helicity-$\pm 1$) and longitudinal (helicity-$0$) 
polarization vectors of the $W^\pm$ bosons are
\begin{eqnarray}
&&\epsilon^{*\mu}_3(\pm )=\mp\frac{1}{\sqrt{2}}
      \left(0,\cos\theta,\mp i, -\sin\theta\right),\qquad
  \epsilon^{*\mu}_4(\pm )=\mp\frac{1}{\sqrt{2}}
      \left(0,-\cos\theta,\mp i, \sin\theta\right),\nonumber\\
&&\epsilon^{*\mu}_3(0)=\frac{\sqrt{\hat{s}}}{2m_W}
      \left(\hat{\beta},\sin\theta,0,\cos\theta\right),\qquad
\epsilon^{*\mu}_4(0)=\frac{\sqrt{\hat{s}}}{2m_W}
      \left(\hat{\beta},-\sin\theta,0,-\cos\theta\right),
\end{eqnarray}
respectively.

\begin{itemize}
\item[{ }] \mbox{ }\hskip 2.5cm TABLE III. Explicit form of 
           the $d$ functions needed.
\end{itemize}

\begin{center}
\begin{tabular}{l}\hline\hline
\mbox{ }\hskip 0.2cm $d^{2}_{2,2}(\theta)=d^{2}_{-2,-2}(\theta)
               =\frac{1}{4}(1+\cos\theta)^2$\mbox{ }\hskip 0.2cm \\
\mbox{ }\hskip 0.2cm $d^{2}_{2,-2}(\theta)=d^{2}_{-2,2}(\theta)
               =\frac{1}{4}(1-\cos\theta)^2$\mbox{ }\hskip 0.2cm \\
\mbox{ }\hskip 0.2cm $d^{2}_{2,1}(\theta)=-d^{2}_{-2,-1}(\theta)
         =-\frac{1}{2}(1+\cos\theta)\sin\theta$\mbox{ }\hskip 0.2cm \\
\mbox{ }\hskip 0.2cm $d^{2}_{2,-1}(\theta)=-d^{2}_{-2,1}(\theta)
         =\frac{1}{2}(1-\cos\theta)\sin\theta$\mbox{ }\hskip 0.2cm \\
\mbox{ }\hskip 0.2cm $d^{2}_{2,0}(\theta)=d^{2}_{-2,0}(\theta)=
 d^{2}_{0,2}(\theta)=d^{2}_{0,-2}(\theta)
               =\sqrt{\frac{3}{8}}\sin^2\theta$\mbox{ }\hskip 0.2cm \\
\mbox{ }\hskip 0.2cm $d^{1}_{0,1}(\theta)=d^{1}_{0,-1}(\theta)
               =\sqrt{\frac{1}{2}}\sin\theta$\mbox{ }\hskip 0.2cm \\
\mbox{ }\hskip 0.2cm $d^{0}_{0,0}(\theta)=1$\mbox{ }\hskip 0.2cm \\
\hline\hline
\end{tabular}
\end{center}
\vskip 0.3cm

The helicity amplitudes can then be parametrized as
\begin{eqnarray}
{\cal M}_{\lambda_1\lambda_2;\lambda_3\lambda_4}(\theta)
 =
e^2\tilde{\cal M}_{\lambda_1\lambda_2;\lambda_3\lambda_4}(\theta)
d^{J_0}_{\Delta\lambda_{12},\Delta\lambda_{34}},
\end{eqnarray}
where $\Delta\lambda_{12}=\lambda_1-\lambda_2$, 
$\Delta\lambda_{34}=\lambda_3-\lambda_4$,
$J_0={\rm max}(|\Delta\lambda_{12}|, |\Delta\lambda_{34}|)$,
and $d^{J_0}_{\Delta\lambda_{12},\Delta\lambda_{34}}$ is the
$d$ function. The explicit form of the $d$ functions needed here 
is listed in Table~3.

We separate the amplitude into the SM contribution and the
new CP-odd contributions with the factor $i$ extracted 
\begin{eqnarray}
\tilde{\cal M}=\tilde{\cal M}_{SM}+i\tilde{\cal M}_{N},
\label{extract}
\end{eqnarray}
where the new contribution can be decomposed in the form   
\begin{eqnarray}
\tilde{\cal M}_{N}=Y_1\tilde{\cal M}^{Y_1}+Y_2\tilde{\cal M}^{Y_2}
                  +Y_3\tilde{\cal M}^{Y_3}+Y_4\tilde{\cal M}^{Y_4}.
\end{eqnarray}
Here we retain only those terms with one insertion of CP-odd operators.

\subsection{The Standard  Model amplitudes}

The process $\gamma\gamma\rightarrow W^+W^-$ is P and CP preserving 
in the SM at tree level.  This leads to the following relations
\begin{eqnarray}
{\rm P}: \ \ \tilde{\cal M}_{\lambda_1\lambda_2;\lambda_3\lambda_4}
   =\tilde{\cal M}_{-\lambda_1,-\lambda_2;-\lambda_3,-\lambda_4},\qquad
{\rm CP}: \ \ \tilde{\cal M}_{\lambda_1\lambda_2;\lambda_3\lambda_4}
   =\tilde{\cal M}_{-\lambda_2,-\lambda_1;-\lambda_4,-\lambda_3}.
\label{P and CP}
\end{eqnarray}
The Bose symmetry leads to the relation;
\begin{eqnarray}
\tilde{\cal M}_{\lambda_1\lambda_2;\lambda_3\lambda_4}
     =\tilde{\cal M}_{\lambda_2\lambda_1;\lambda_3\lambda_4}\ \
    (\cos\theta \rightarrow -\cos\theta).
\label{Bose}
\end{eqnarray}
Let us rewrite the amplitude in the form 
\begin{eqnarray}
\tilde{\cal M}_{SM}=\frac{\tilde{\cal N}^{SM}}{1
                    -\hat{\beta}^2\cos^2\theta},
\end{eqnarray}
by extracting the $t$- and $u$-channel $W$ boson propagator factor.
It is clear that the coefficients $\tilde{\cal N}$'s satisfy the same 
P, CP and Bose-symmetry relations as $\hat{{\cal M}}_{SM}$.
We find for the positive photon helicity ($\lambda_1=+$), 
\begin{eqnarray}
\begin{array}{ll}
\tilde{\cal N}^{SM}_{++;++}=2(1+\hat{\beta})^2,\qquad &
\tilde{\cal N}^{SM}_{++;+0}=\tilde{\cal N}^{SM}_{++;+-}
                      =\tilde{\cal N}^{SM}_{++;0+}=0, \\
\tilde{\cal N}^{SM}_{++;00}=-\frac{8}{\hat{r}}, \qquad &
\tilde{\cal N}^{SM}_{++;0-}=\tilde{\cal N}^{SM}_{++;-+}
                      =\tilde{\cal N}^{SM}_{++;-0}=0, \\
\tilde{\cal N}^{SM}_{++;--}=2(1-\hat{\beta})^2,\qquad &
\tilde{\cal N}^{SM}_{+-;++}=\frac{32}{\sqrt{6}\hat{r}},\qquad
\hskip 0.3cm \tilde{\cal N}^{SM}_{+-;+-}=8,\\
\tilde{\cal N}^{SM}_{+-;+0}=\tilde{\cal N}^{SM}_{+-;0+}
                      =\frac{8}{\sqrt{2\hat{r}}},\qquad  &
\tilde{\cal N}^{SM}_{+-;00}=4\sqrt{\frac{2}{3}}(2-\hat{\beta}^2),
\end{array}
\end{eqnarray}
where $\hat{r}=\hat{s}/m^2_W$. 
The other remaining coefficients can be obtained by using  
the P and CP relations (\ref{P and CP}) and the Bose-symmetry
We note the following three features of the SM amplitudes;
\begin{itemize}
\item The amplitudes for producing two $W$'s with the non-vanishing
      total spin component along the $W$ boson momentum direction 
      ($\Delta\lambda_{34}$) vanish when the initial state has
      $J_z=\Delta\lambda_{12}=0$.
\item The amplitude for producing two longitudinal $W$'s from a $J_z=0$
      initial state is suppressed by a factor of $1/\hat{r}$
      in the SM. The same behaviour should appear in the production
      of two charged scalars such as 
      $\gamma\gamma\rightarrow \pi^+\pi^-$.
\item The amplitudes for producing two right-(left-)handed $W$'s 
      from two left-(right-)handed photons is suppressed by a factor
      of $1/\hat{r}^2$.  
\end{itemize}
The results are consistent with those by Yehudai\cite{ScFs} and by 
B\'{e}langer, {\it et.al}\cite{GbGc}.

\subsection{CP-odd amplitudes}

Every CP-odd amplitude and its CP-conjugate amplitude satisfies the 
following relation
\begin{eqnarray}
\tilde{\cal M}^{Y_i}_{\lambda_1\lambda_2;\lambda_3\lambda_4}
           =
 -\tilde{\cal M}^{Y_i}_{-\lambda_2,-\lambda_1;-\lambda_4,-\lambda_3}
  \qquad (i=1,2,3,4),
\end{eqnarray}
since the factor of $i$ is extracted in the full helicity amplitude 
(\ref{extract}).
It then follows that any CP self-conjugate amplitude has vanishing 
contribution from the CP-odd terms;
\begin{eqnarray}
\tilde{\cal M}^{Y_i}(\pm\mp;\pm\mp)=\tilde{\cal M}^{Y_i}(\pm\mp;\mp\pm)
     =\tilde{\cal M}^{Y_i}(\pm\mp;00)=0
   \qquad (i=1,2,3,4).
\end{eqnarray}

The $Y_1$ and $Y_2$ terms contribute to the $t$- and $u$- channels 
and $Y_2$ contributes to the contact $\gamma\gamma WW$ diagram as well.
By using the notation
\begin{eqnarray}
\tilde{\cal M}^{Y_i}
   =\frac{\tilde{\cal N}^{Y_i}}{1-\hat{\beta}^2\cos^2\theta},
    \qquad (i=1,2),
\end{eqnarray}
we find that the non-vanishing $Y_1$ contributions
are  
\begin{eqnarray}
\tilde{\cal N}^{Y_1}_{++;++}&=&-\tilde{\cal N}^{Y_1}_{--;--}=
  8\bigg[2+\hat{\beta}(1+\cos^2\theta)\bigg],\nonumber\\
\tilde{\cal N}^{Y_1}_{++;+0}&=&-\tilde{\cal N}^{Y_1}_{--;0-}=
\tilde{\cal N}^{Y_1}_{++;0+} = -\tilde{\cal N}^{Y_1}_{--;-0}=
  4\sqrt{\hat{r}}\hat{\beta}(1+\hat{\beta}),\nonumber\\
\tilde{\cal N}^{Y_1}_{++;00}&=&-\tilde{\cal N}^{Y_1}_{--;00}=
  -4\hat{r}(1-\hat{\beta}^2\cos^2\theta),\nonumber\\
\tilde{\cal N}^{Y_1}_{++;0-}&=&-\tilde{\cal N}^{Y_1}_{--;+0}=
\tilde{\cal N}^{Y_1}_{++;-0} = -\tilde{\cal N}^{Y_1}_{--;0+}=
  -4\sqrt{\hat{r}}\hat{\beta}(1-\hat{\beta}),\nonumber\\
\tilde{\cal N}^{Y_1}_{++;--}&=&-\tilde{\cal N}^{Y_1}_{--;++}=
  8\bigg[2-\hat{\beta}(1+\cos^2\theta)\bigg],\nonumber\\
\tilde{\cal N}^{Y_1}_{+-;++}&=&-\tilde{\cal N}^{Y_1}_{+-;--}=
\tilde{\cal N}^{Y_1}_{-+;++} = -\tilde{\cal N}^{Y_1}_{-+;--}=
  -\frac{32}{\sqrt{6}}\hat{\beta},\nonumber\\
\tilde{\cal N}^{Y_1}_{+-;+0}&=&-\tilde{\cal N}^{Y_1}_{+-;0-}=
\tilde{\cal N}^{Y_1}_{+-;0+} = -\tilde{\cal N}^{Y_1}_{+-;-0}
                        =  \tilde{\cal N}^{Y_1}_{-+;+0}\nonumber\\
      &=&-\tilde{\cal N}^{Y_1}_{-+;0-}=
\tilde{\cal N}^{Y_1}_{-+;0+} = -\tilde{\cal N}^{Y_1}_{-+;-0}=
  -4\sqrt{2\hat{r}}\hat{\beta},
\label{Amp1}
\end{eqnarray}
and the non-vanishing $Y_2$ contributions are 
\begin{eqnarray}
\tilde{\cal N}^{Y_2}_{++;++}&=&-\tilde{\cal N}^{Y_2}_{--;--}=
  -12\hat{r}\bigg[1-3\hat{\beta}+\hat{\beta}^2+\hat{\beta}^3
  +(1+\hat{\beta}-3\hat{\beta}^2+\hat{\beta}^3)\cos^2\theta\bigg],
   \nonumber\\
\tilde{\cal N}^{Y_2}_{++;+0}&=&-\tilde{\cal N}^{Y_2}_{--;0-}=
\tilde{\cal N}^{Y_2}_{++;0+} = -\tilde{\cal N}^{Y_2}_{--;-0}=
  -24\sqrt{\hat{r}}(\hat{\beta}+1)(\hat{\beta}-2)\cos\theta,
    \nonumber\\
\tilde{\cal N}^{Y_2}_{++;+-}&=&-\tilde{\cal N}^{Y_2}_{--;+-}=
\tilde{\cal N}^{Y_2}_{++;-+} = -\tilde{\cal N}^{Y_2}_{--;-+}=
  -\frac{48}{\sqrt{6}}\hat{r}(1+\hat{\beta}^2),\nonumber\\
\tilde{\cal N}^{Y_2}_{++;00}&=&-\tilde{\cal N}^{Y_2}_{--;00}=
  96\sin^2\theta,\nonumber\\
\tilde{\cal N}^{Y_2}_{++;0-}&=&-\tilde{\cal N}^{Y_2}_{--;+0}=
\tilde{\cal N}^{Y_2}_{++;-0} = -\tilde{\cal N}^{Y_2}_{--;0+}=
  24\sqrt{\hat{r}}(\hat{\beta}-1)(\hat{\beta}+2)\cos\theta,\nonumber\\
\tilde{\cal N}^{Y_2}_{++;--}&=&-\tilde{\cal N}^{Y_2}_{--;++}=
  -12\hat{r}\bigg[1+3\hat{\beta}+\hat{\beta}^2-\hat{\beta}^3
  +(1-\hat{\beta}-3\hat{\beta}^2-\hat{\beta}^3)\cos^2\theta\bigg],
  \nonumber\\
\tilde{\cal N}^{Y_2}_{+-;++}&=&-\tilde{\cal N}^{Y_2}_{+-;--}=
\tilde{\cal N}^{Y_2}_{-+;++} = -\tilde{\cal N}^{Y_2}_{-+;--}=
  \frac{48}{\sqrt{6}}\hat{r}(1+\hat{\beta}^2),\nonumber\\
\tilde{\cal N}^{Y_2}_{+-;+0}&=&-\tilde{\cal N}^{Y_2}_{+-;0-}=
\tilde{\cal N}^{Y_2}_{+-;0+} = -\tilde{\cal N}^{Y_2}_{+-;-0}
                             =  \tilde{\cal N}^{Y_2}_{-+;+0}\nonumber\\
      &=&-\tilde{\cal N}^{Y_2}_{-+;0-}=
\tilde{\cal N}^{Y_2}_{-+;0+} = -\tilde{\cal N}^{Y_2}_{-+;-0}=
  24\sqrt{2\hat{r}}\hat{\beta}.
\label{Amp2}
\end{eqnarray}

The two contributions behave differently at high energies.
The $Y_1$ contributions are dominant in the amplitudes for
producing two longitudinal $W$'s from the $J_Z=0$ initial photon
state 
\begin{eqnarray}
\tilde{\cal N}^{Y_1}_{\pm\pm;00}\rightarrow \mp 4\hat{r}\sin^2\theta,
\end{eqnarray}
while the $Y_2$ contributions are dominant in the amplitudes for
producing two transverse $W$'s except for the $(\pm\pm;\pm\pm)$
modes 
\begin{eqnarray}
\tilde{\cal N}^{Y_2}_{\pm\pm;\mp\mp}&\rightarrow & 
                     \mp 48\hat{r}\sin^2\theta,\nonumber\\
\tilde{\cal N}^{Y_2}_{\pm\pm;+-}
 =\tilde{\cal N}^{Y_2}_{\pm\pm;-+}&\rightarrow  &
                     \mp 16\sqrt{6}\hat{r},\nonumber\\ 
\tilde{\cal N}^{Y_2}_{+-;\pm\pm}
 =\tilde{\cal N}^{Y_2}_{-+;\pm\pm}&\rightarrow &
                     \pm 16\sqrt{6}\hat{r}.
\end{eqnarray}
The high-energy behavior of two sets of amplitudes (\ref{Amp1})
and (\ref{Amp2}) are in sharp 
contrast to that of the SM amplitudes whose dominant contributions
are in the $(\pm\pm;\pm\pm)$, $(\pm\mp;\pm\mp)$, 
and $(\pm\mp;00)$ modes. Because of this, interference between
different helicity amplitudes are essential to observe significant
CP violation effects. Use of the linearly polarized photon beams
allow us to study interference between the leading CP-even (SM) 
amplitudes and the leading CP-odd amplitudes.
In our approximation of neglecting the one-loop $\gamma\gamma H$ 
vertex of the SM, there is no contribution from the $Y_3$ term;
\begin{eqnarray}
\tilde{\cal M}^{Y_3}=0.
\end{eqnarray}

On the other hand $Y_4$ contributes to the $s$-channel
scalar exchange diagram in the helicity amplitudes with 
$\Delta\lambda_{12}=\Delta\lambda_{34}=0$. 
An explicit calculation shows that the non-vanishing amplitudes, 
$\tilde{\cal M}_{Y_4}$, are as follows;
\begin{eqnarray}
&&\tilde{\cal M}^{Y_4}_{++;++}=-\tilde{\cal M}^{Y_4}_{--;--}
       =4\chi_H(\hat{s}),\nonumber\\
&&\tilde{\cal M}^{Y_4}_{++;00}=-\tilde{\cal M}^{Y_4}_{--;00}
       =-\hat{r}
         (1+\hat{\beta}^2)\chi_H(\hat{s}),\nonumber\\
&&\tilde{\cal M}^{Y_4}_{++;--}=-\tilde{\cal M}^{Y_4}_{--;++}
       =4\chi_H(\hat{s}),
\end{eqnarray}
where $\chi_H$ is the Higgs propagator factor
\begin{eqnarray}
\chi_H(\hat{s})=\frac{\hat{s}}{\hat{s}-m^2_H+im_H\Gamma_H}.
\end{eqnarray}
In the subsequent numerical studies, we examine the case with 
$m_H=100$ GeV where the width $\Gamma_H$ is safely neglected.
We will study the $m_H\geq 2m_W$ case elsewhere, since there
both the tree- and one-loop SM amplitudes are relevant.
\vskip 1.5cm

\section{Differential cross section}

In counting experiments where the final $W$ polarizations are not
analyzed, we measure only the following combinations:
\begin{eqnarray}
\sum_X M_{\lambda_1\lambda_2}M^*_{\lambda^\prime_1\lambda^\prime_2}
 =
e^4\sum_{\lambda_3}\sum_{\lambda_4}
   \tilde{\cal M}_{\lambda_1\lambda_2;\lambda_3\lambda_4}
   \tilde{\cal M}^*_{\lambda^\prime_1\lambda^\prime_2;\lambda_3\lambda_4}.
\label{distr5}
\end{eqnarray}
We then find $\Sigma_{\rm unpol}$, $ \Sigma_{02}$, $\Delta_{02}$,
$\Sigma_{22}$, and  $\Sigma_{00}$ from Eq.~(13).
The differential cross section for a fixed angle $\chi$ is 
\begin{eqnarray}
\frac{{\rm d}^2\sigma}{{\rm d}\cos\theta{\rm d}\phi}(\chi)
&=&\frac{\alpha^2}{8\hat{s}(1-\hat{\beta}^2\cos^2\theta)^2}
 \Bigg\{\hat{\Sigma}_{\rm unpol}-\frac{1}{2}Re\bigg[
 \left(\eta{\rm e}^{-i(\chi+\phi)}
 +\bar{\eta}{\rm e}^{-i(\chi-\phi)}\right)
 \hat{\Sigma}_{02}\bigg]\nonumber\\
&+&\frac{1}{2}Re\bigg[
  \left(\eta{\rm e}^{-i(\chi+\phi)}
  -\bar{\eta}{\rm e}^{-i(\chi-\phi)}\right)
  \hat{\Delta}_{02}\bigg]+\eta\bar{\eta}Re\bigg[
  {\rm e}^{-2i\phi}\hat{\Sigma}_{22}+{\rm e}^{-2i\chi}\hat{\Sigma}_{00}
  \bigg]\Bigg\},\\
\Sigma_i&=&\frac{e^2\hat{\Sigma}_i}{(1-\hat{\beta}^2\cos^2\theta)^2},
   \qquad
\Delta_{02}=\frac{e^2\hat{\Delta}_{02}}{(1-\hat{\beta}^2\cos^2\theta)^2},
\end{eqnarray}
for $i={\rm unpol},{02},{22}$, and ${00}$.

We first note that all the real parts of the distributions (\ref{distr5}) 
are independent of the anomalous CP-odd form factors $Y_i$ 
up to linear order
\begin{eqnarray}
&&\hat{\Sigma}_{\rm unpol}=38-4\hat{\beta}^2(3-8\cos^2\theta)
                  +6\hat{\beta}^4(1+\sin^4\theta),\nonumber\\
&&\begin{array}{ll}
  {\cal R}(\hat{\Sigma}_{02})
         =\frac{96}{\hat{r}}\hat{\beta}^2\sin^2\theta,\ \ &
  {\cal R}(\hat{\Delta}_{00})=0,\\
  {\cal R}(\hat{\Sigma}_{22})=6\hat{\beta}^4\sin^4\theta,\ \ &
  {\cal R}(\hat{\Sigma}_{00})=\frac{96}{\hat{r}^2}.
  \end{array} 
\end{eqnarray}
On the other hand, two CP-odd distributions, 
${\cal I}(\hat{\Sigma}_{02})$ and ${\cal I}(\hat{\Sigma}_{00})$, 
have contributions from the $Y_1$, $Y_2$, and $Y_4$ terms
\begin{eqnarray}
{\cal I}(\hat{\Sigma}_{02})&=&-4\hat{r}\hat{\beta}^2
       \bigg[4(1-\hat{\beta}^2\cos^2\theta)R(Y_1)
            +48(3+\hat{\beta}^2\cos^2\theta)R(Y_2)\nonumber\\
           &&{ } \hskip 2.7cm 
            +(5-3\hat{\beta}^2)(1-\hat{\beta}^2\cos^2\theta)
             R(Y_4\chi_H)\bigg]\sin^2\theta,\\
{\cal I}(\hat{\Sigma}_{00})&=&24
       \bigg[4R(Y_1)-4\hat{r}(1+3\hat{\beta}^2)R(Y_2)
            +(1+\hat{\beta}^2)R(Y_4\chi_H)\bigg]
             (1-\hat{\beta}^2\cos^2\theta).
\label{Main1}
\end{eqnarray}

A few comments on the CP-odd distributions are in order.
\begin{itemize}
\item ${\cal I}(\hat{\Sigma}_{02})$ has $\hat{\beta}^2$ as an 
      overall factor such that the contribution vanishes at the threshold, 
      whereas ${\cal I}(\hat{\Sigma}_{00})$ does not. 
\item Both CP-odd distributions have the angular terms ($\sin^2\theta$
      and $1-\hat{\beta}^2\cos^2\theta$) which become largest
      at the scattering angle $\theta=\pi/2$, where the SM contributions 
      are generally small.  We, therefore, expect large CP-odd 
      asymmetries at $\theta\approx\pi/2$.
\item Each term in ${\cal I}(\Sigma_{02})$ has a different angular
      dependence which allows us to disentangle them. On the other hand, 
      we note that all the terms in ${\cal I}(\Sigma_{00})$ mode all the
      contributions have the same angular dependence. The only way
      to distinguish them is to study its energy dependence. 
      We show that this can be efficiently done by adjusting 
      the laser beam frequency in the Compton backscattering mode.
\item At high energies ($\hat{r}>>1$), $R(Y_1)$ and $R(Y_3)$ are measured
      from ${\cal I}(\Sigma_{02})$, whereas $R(Y_2)$ affects both
      ${\cal I}(\Sigma_{00})$ and ${\cal I}(\Sigma_{02})$.
\end{itemize}
\vskip 1.5cm

\section{Observable consequences of CP-odd couplings}
\label{sec:observable}

Let us estimate the various experimental branching fractions
of $W$ decays.
Consider the decay of each $W$ into a 
fermion-antifermion pair (quark-antiquark $q_1\bar{q}_2$ or
charged lepton-neutrino $l\nu_l$) at tree level. 
The branching ratio for $W^-\rightarrow l\bar{\nu}_l$ 
($l=e,\mu$, or $\tau$) is about 10\% each\cite{PDG94}. 
We thus expect the following final state combinations:
\begin{eqnarray}
\begin{array}{ll}
 (q\bar{q})(q\bar{q})\Rightarrow 4{\rm jets}\hskip 0.5cm  
& 49\%,\\
 (q\bar{q})(l\nu)\Rightarrow {\rm dijet}+l^\pm+\not\!{p}\hskip 0.5cm 
& 42\%,\\
 (l\bar{\nu})(\bar{l}\nu)\Rightarrow l^+l^-+\not\!{p}\hskip 0.5cm
& \ \ 9\%,
\end{array}
\end{eqnarray}
where $\not\!\!{p}$ stands for the momentum of the escaping neutrino(s).
The dijet$+l^\pm$ mode is most amenable for $W$-spin analysis. 
In our analysis, no spin analysis for the decaying $W$'s is 
required. In case of ${\cal I}(\Sigma_{00})$, not even the scattering
plane needs to be identified. Even if one excludes the $\tau^+\tau^-+
\not\!{p}$ modes of 1\%, the remaining 99\% of the events can be used
to measure ${\cal I}(\Sigma_{00})$. On the other hand, the scattering
plane should be identified to measure ${\cal I}(\Sigma_{02})$. It
is worth noting that the charge of the decaying $W$ is not needed to
extract ${\cal I}(\Sigma_{02})$. Therefore all the modes except for
the $l^+l^-+\not\!{p}$ modes (9\%) can be used 
for ${\cal I}(\Sigma_{02})$.

The $\gamma\gamma\rightarrow W^+W^-$ reaction has a much 
larger cross section than heavy fermion-pair production such as 
$\gamma\gamma\rightarrow t\bar{t}$ and, furthermore, the total cross 
section approaches a constant value at high c.m. energies. 
At $\sqrt{\hat{s}}=500$ GeV the total cross section is about 80~pb, 
while the $t\bar{t}$ cross section is about 1~pb. So there exist no
severe background problems. 
In the following analysis we simply assume that all the $W$ pair events 
can be used. It would be rather straightforward to include
the effects from any experimental cuts and efficiencies in addition to
the branching factors discussed above. 

We present our numerical analysis at the following set of collider
parameters:		
\begin{eqnarray}
\sqrt{s}=0.5\ \ {\rm and}\ \  1.0\ \ {\rm TeV},\qquad 
\kappa^2\cdot L_{ee}=20\ \ {\rm fb}^{-1}.
\label{parameters}
\end{eqnarray}
The dimensionless parameter $x$, which is dependent on the laser 
frequency $\omega_0$, is treated as an adjustable parameter.
We note that $\kappa=1$ is the maximally allowed value for the
$e$-$\gamma$ conversion coefficient $\kappa$ and it may be as small
as $\kappa=0.1$ if the collider is optimized for the $e^+e^-$ 
model\cite{GKS}. All one should note is that the significance
of the signal scales as $(\epsilon\cdot\kappa^2\cdot L_{ee})$,
where $\epsilon$ denotes the overall detection efficiency that
is different for $A_{00}$ and $A_{02}$.

\subsection{Statistical significance of possible signals}

The two CP-odd integrated asymmetries, $A_{00}$ and $A_{02}$,
depend linearly on the form factors, $R(Y_1)$, $R(Y_2)$, and $R(Y_4)$ 
in the approximation that only the terms linear in the
form factors are retained.  We present the sensitivities 
to each form factor, assuming that the other
form factors are zero. The analyses are catalogued into 
two parts: the $\gamma(\gamma) WW$ part and the $\gamma\gamma H$ part.

Folding the photon luminosity spectrum and integrating the 
distributions over the polar angle $\theta$, we obtain the 
$x$-dependence of available event rates:
\begin{eqnarray}
  \left(\begin{array}{c}
      N_{\rm unpol} \\
      N_{02}\\
      N_{00}\end{array}\right)
    =\kappa^2L_{ee}\frac{\pi\alpha^2}{2s}
     \int^{\tau_{\rm max}}_{\tau_{\rm min}}\frac{{\rm d}\tau}{\tau}
     \int^1_{-1} {\rm d}\cos\theta 
     \frac{\hat{\beta}
       \langle\phi_0\phi_0\rangle_\tau}{(1-\hat{\beta}^2\cos^2\theta)^2}
   \left(\begin{array}{c}
     \hat{\Sigma}_{\rm unpol}\\
     {\cal A}_\eta{\cal I}\left(\hat{\Sigma}_{02}\right)\\
     {\cal A}_{\eta\eta}{\cal I}\left(\hat{\Sigma}_{00}\right)
       \end{array}\right),
\end{eqnarray}
where $\tau_{max}=(x/(1+x))^2$ and $\tau_{min}=4m^2_W/s$.
One measure of the significance of a CP-odd asymmetry
is the standard deviation ${\cal N}^a_{SD}$ by which
the asymmetry exceeds the expected statistical fluctuation of
the background distribution; for $a=02$ and $00$
\begin{eqnarray}
{\cal N}^a_{SD}=\frac{|A_a|}{\sqrt{2/\epsilon N_{\rm unpol}}}.
\end{eqnarray}
Here $\epsilon$ is for the sum of $W$ branching
fractions available, which is taken to be
\begin{eqnarray}
\varepsilon=
\left\{\begin{array}{cl}
       100\% \hskip 0.3cm \ \ & {\rm for}\ \  N_{00},\\
       91\%  \hskip 0.3cm \ \ & {\rm for}\ \  N_{02}.
       \end{array}\right.
\end{eqnarray}
Separating the asymmetry $A_a$ into four independent parts as
\begin{eqnarray}
A_a=R(Y_1)A^{Y_1}_a+R(Y_2)A^{Y_2}_a+R(Y_4)A^{Y_4}_a,
\end{eqnarray}
and considering each form factor separately,
we obtain the $1$-$\sigma$ allowed upper bounds of the form factors
($i=1,2,4$)
\begin{eqnarray}
{\rm Max}(|R(Y_i)|_a)
   =\frac{\sqrt{2}}{|A^{Y_i}_a\sqrt{\epsilon N_{\rm unpol}}|}, 
\end{eqnarray}
if no asymmetry is found. The $N_{SD}$-$\sigma$ upper bound  
is determined simply by multiplying 
${\rm Max}(|R(Y_i)|_a)$ and ${\rm Max}(|I(Y_4)|_a)$ by $N_{SD}$.

\subsection{The $\gamma WW$ and $\gamma\gamma WW$ vertices: 
            $Y_1$ and $Y_2$}

The parity-violating form factors $Y_1$ and $Y_2$ respect charge
conjugation invariance and they are related to the $W$ electric dipole
moment(EDM) $d_W$ and the $W$ magnetic quadrupole moment(MQD)
$\tilde{Q}_W$ of $W^+$ by
\begin{eqnarray}
d_W=\frac{2e}{m_W}\left(Y_1+6Y_2\right),\qquad 
\tilde{Q}_W=-\frac{4e}{m^2_W}\left(Y_1-6Y_2\right).
\end{eqnarray}
There are strong indirect phenomenological constraints on the above 
couplings arising from the EDM of the electron and neutron\cite{WmAq}. 
However, we should note that there is a possibility of cancellation 
among different contributions which renders these indirect constraints 
ineffective. 
Direct studies of $W$-pair production at high energies are quite 
complementary to the precision experiments at low energies. 
Although the interplay between high- and low-energy experimental 
constraints is important, the latter constraints can not replace the 
role of high-energy experiments.

Figs.4(a) and (b) show the $x$ dependence of the sensitivities 
to $R(Y_1)$, which are obtained from $A_{02}$ and $A_{00}$, 
respectively, for $\sqrt{s}=0.5$ TeV and $\sqrt{s}=1$ TeV. 
The $x$ dependence of the sensitivities to $R(Y_2)$ are shown 
in Figs.~5(a) and (b). In both figures, the solid lines are for
$\sqrt{s}=0.5$ TeV and the long-dashed lines for $\sqrt{s}=1.0$ TeV.
Let us make a few comments on the results shown in the two 
figures (Figs.~4 and 5) and Table~4.
\begin{itemize}
\item The sensitivities, especially from the asymmetry $A_{00}$ mode,
      depend strongly on the value of $x$. For
      smaller $x$ values are favored for $A_{00}$, while
      relatively large $x$ values favored for $A_{02}$.
      This property can be clearly understood by noting that
      $\hat{A}_{00}$ gets suppressed as the $\gamma\gamma$
      c.m. energy increases, while $\hat{A}_{02}$ does not.
\item The optimal sensitivities on $R(Y_2)$ are very much improved 
      as the $e^+e^-$ c.m. energy increases from $0.5$ TeV to 
      1 TeV while those of $R(Y_1)$ are a little improved.
      The optimal $x$ values are reduced as the c.m. energy increases. 
\item At the two $\sqrt{s}$ values the asymmetries $A^{Y_1}_{00}$
      gives stronger sensitivities than $A^{Y_1}_{02}$ to $R(Y_1)$, 
      while the two symmetries $A^{Y_2}_{02}$ and $A^{Y_2}_{00}$ 
      gives rather similar sensitivities to $R(Y_2)$.
      These properties can be understood from the 
      $\hat{s}$ dependence of the corresponding CP-odd 
      distributions (\ref{Main1}).
\end{itemize}
\begin{enumerate}
\item[{ }] TABLE IV. The best $1$-$\sigma$ bounds of 
          the CP-odd form factors, $R(Y_1)$ and $R(Y_2)$, and 
          their corresponding  
          $x$ values for $\sqrt{s}=0.5$ and $1$ TeV.
\end{enumerate}

\begin{center}
\begin{tabular}{|c||c|c||c|c|}\hline\hline
       { }            &  \multicolumn{2}{c||}{$A_{02}$}  
                      &  \multicolumn{2}{c|}{$A_{00}$}  \\
\hline
    $\sqrt{s}$ (TeV)  & $0.5$  &  $1.0$  
                      & $0.5$  &  $1.0$ \\
\hline\hline
       $x$            & $1.83$ &  $0.96$
                      & $0.75$ &  $0.31$\\
\hline
  ${\rm Max}(|R(Y_1)|)$  
      &\mbox{ }\hskip 0.2cm  $1.1\times 10^{-2}$\mbox{ }\hskip 0.2cm  
      &\mbox{ }\hskip 0.2cm  $5.0\times 10^{-3}$\mbox{ }\hskip 0.2cm 
      &\mbox{ }\hskip 0.2cm  $3.2\times 10^{-3}$\mbox{ }\hskip 0.2cm  
      &\mbox{ }\hskip 0.2cm  $2.2\times 10^{-3}$\mbox{ }\hskip 0.2cm \\
\hline\hline
       $x$            & $2.09$ &  $1.23$
                      & $1.11$ &  $0.59$\\
\hline
  ${\rm Max}(|R(Y_2)|)$  
      &\mbox{ }\hskip 0.2cm $2.4\times 10^{-4}$\mbox{ }\hskip 0.2cm 
      &\mbox{ }\hskip 0.2cm $9.0\times 10^{-5}$\mbox{ }\hskip 0.2cm
      &\mbox{ }\hskip 0.2cm $2.6\times 10^{-4}$\mbox{ }\hskip 0.2cm 
      &\mbox{ }\hskip 0.2cm $1.1\times 10^{-4}$\mbox{ }\hskip 0.2cm\\
\hline\hline
\end{tabular}
\end{center}
\vskip 0.3cm

The above results underlie the importance of having adjustable 
laser frequencies, which allows us to select the regime where
 each contribution becomes dominant. 
We find that the two-photon mode allows us to reach
the limit that $R(Y_1)$ is of the order of $10^{-3}$ and
$R(Y_2)$ is of the order of $10^{-4}$ or less.

\subsection{The $\gamma\gamma H$ vertex: $Y_4$}

The $\gamma\gamma H$ vertex $Y_4$ can be studied in the process
$\gamma\gamma\rightarrow H$\cite{JgHh}, where the interference between
the 1-loop SM amplitudes and the new CP-odd amplitudes lead to 
observable CP-odd asymmetries. In this paper, we study the sensitivity
of the process $\gamma\gamma\rightarrow W^+W^-$ to the CP-odd
$\gamma\gamma H$ coupling $Y_4$ where $m_H$ is below the $W$-pair
threshold. For an actual numerical analysis we set $m_H=100$ GeV 
and assume that its width is negligible. Our results are insensitive
to $m_H$ as long as $m_H< 2m_W$.

\begin{enumerate}
\item[{ }] TABLE V. The $1$-$\sigma$ sensitivities 
          to the CP-odd form factor, $R(Y_4)$, and their
          corresponding  $x$ values for $\sqrt{s}=0.5$ and $1$ TeV. 
          Here, $m_H=100$ GeV.
\end{enumerate}

\begin{center}
\begin{tabular}{|c||c|c||c|c|}\hline\hline
       { }            &  \multicolumn{2}{c||}{$A_{02}$}  
                      &  \multicolumn{2}{c|}{$A_{00}$}  \\
\hline
    $\sqrt{s}$ (TeV)  & $0.5$  &  $1.0$  
                      & $0.5$  &  $1.0$ \\
\hline\hline
       $x$            & $1.43$ &  $0.69$
                      & $0.76$ &  $0.31$\\
\hline
  ${\rm Max}(|R(Y_4)|)$  
   &\mbox{ }\hskip 0.2cm  $1.1\times 10^{-2}$\mbox{ }\hskip 0.2cm  
   &\mbox{ }\hskip 0.2cm  $6.4\times 10^{-3}$\mbox{ }\hskip 0.2cm
   &\mbox{ }\hskip 0.2cm  $7.5\times 10^{-3}$\mbox{ }\hskip 0.2cm  
   &\mbox{ }\hskip 0.2cm  $5.0\times 10^{-3}$\mbox{ }\hskip 0.2cm \\
\hline\hline
\end{tabular}
\end{center}
\vskip 0.3cm

The best sensitivities to $R(Y_4)$ from
the asymmetries $A_{02}$ and $A_{00}$ and their corresponding $x$
values for $\sqrt{s}=0.5$ and $1.0$ TeV are listed in Table~5.
Two asymmetries give the approximately same sensitivities
at the same $x$ value. The doubling of the $e^+e^-$ c.m. energy
improves the sensitivity so much and renders the optimal $x$
values smaller than those at $\sqrt{s}=0.5$ TeV. 
Fig.~6 show the very strong $x$ dependence of the $R(Y_4)$ $1$-$\sigma$ 
sensitivities. Quantitatively we find that the constraints on $R(Y_4)$  
are of the order of $10^{-3}$ for $m_H=100$ GeV at $\sqrt{s}=0.5$ and 
$1.0$ TeV.

\subsection{Model expectations}

In order to assess the usefulness of the two-photon mode
with polarized photons it is useful to estimate
the expected size of the CP-odd form factors 
in a few specific models with reasonable physics assumptions. 
Several works\cite{DcWk} have estimated  the size of the $W$ EDM 
in various models beyond the SM. 
They have shown that the $W$ EDM can be of the order 10$^{-20}$ 
$(e{\rm cm})$ in the multi-Higgs-doublet model and the 
supersymmetric SM, corresponding to $Y_1$ and $Y_2$ of the order 
of 10$^{-4}$. It is predicted of about 10$^{-22}$ and less than 
10$^{-38}$ $(e{\rm cm})$ in the left-right model and the SM, 
respectively, 

In more general, if these vertices appear in the one-loop 
level\cite{CaMe} the coefficients $f_i$ may contain a factor 
of $1/16pi^2$. By setting all $f_i$'s to be $1/16\pi^2$ and 
setting $\Lambda=v=246$ GeV, we find
\begin{eqnarray}
|Y_1|\sim |Y_3|\sim |Y_4| \sim 10^{-3},\qquad
|Y_2|\sim 10^{-4}.
\label{limits}
\end{eqnarray}
The above order of magnitude estimates (\ref{limits}) of 
the form factors  are consistent with the values expected in some 
specific models. 

It is worth remarking that the two-photon experiments may allow us to
probe the CP-odd effects of the expected size (\ref{limits}).
The two-photon collider with polarized photons and adjustable laser
frequency can play a crucial role in probing CP violation in 
the bosonic sector.

\subsection{Comparison of the $\gamma\gamma$ mode and the $e^+e^-$ mode}

The initial $e^+e^-$ state of the $e^+e^-\rightarrow W^+W^-$ process
is (almost) CP-even due to the very small electron mass. 
It is then clear that the initial electron beam polarizations
are not so useful to construct large CP-odd asymmetries. 
CP-violating $W$ interactions can be probed only via spin/angular 
correlations of the decaying $W$'s. For $L_{ee}=20$ fb$^{-1}$ 
and $\kappa=1$, we compare the constraints from the two-photon 
mode with those from the $e^+e^-$ mode by studying the $W^\pm$ decay
correlations at $\sqrt{s}=0.5$ TeV.

The process $e^+e^-\rightarrow W^+W^-$\cite{KhRp,PkPm} has been 
investigated in detail. For the present comparison let us  
refer to the work by Kalyniak, {\it et.al}\cite{PkPm}, where 
they have assumed $L_{ee}=50$ fb$^{-1}$ and a perfect detector.
Readjusting the $e^+e^-$ integrated luminosity to $20$ fb$^{-1}$,
we can summarize their findings; the total cross section with the pure
leptonic decay modes of the $W$'s gives the constraint
$|R(Y_1)|\leq 5\times 10^{-2}$. 
below the expected level of statistical precision of approximately
The two-photon mode is much more promising than the $e^+e^-$ mode 
in probing CP violation in the $W$ pair production, if $\kappa\sim 1$
$e\gamma$ conversion rate is technically achieved. 
\vskip 1.5cm

\section{Summary and conclusions}
\label{sec:conclusion}

In this paper we have made a systematic study of observable 
asymmetries related with two polarized-photon collisions via
the Compton backscattered laser beam at future linear colliders,
which could serve as tests of possible CP-violating effects.
We have described in a general framework how photon polarization
is employed to study CP invariance in the initial two-photon state.
We have considered the most general dimension-six CP-odd operators 
in the scalar and vector boson sector, preserving all the SM gauge 
symmetries in the linear realization of the electroweak symmetry
breaking.

Limiting ourselves to purely linearly-polarized photon beams, 
we have constructed two CP-odd asymmetries in the process
$\gamma\gamma\rightarrow W^+W^-$. The CP-odd asymmetries can be
extracted by simply adjusting the angle between the polarization
vectors of two laser beams. 
We have found that the sensitivities of the CP-odd asymmetries to
the CP-odd form factors depend strongly on the $e^+e^-$ c.m. energy
and the laser beam frequency.

In Tables~4 and 5 the maximal sensitivities of the CP-odd form factors 
and the corresponding $x$ values have been shown for $\sqrt{s}=0.5$ 
and $1$ TeV with $\kappa^2\cdot L_{ee}=20$ fb$^{-1}$.
The sensitivities are high enough to probe CP-odd new interactions beyond
the limits from some specific models with resonable physics assumptions.

We have found that, for $\kappa\sim 1$, a counting experiment in the 
two-photon mode with adjustable laser frequency 
can give much stronger constraints on the $W$ EDM and magnetic quadratic 
moment (MQD) than the $e^+e^-$ mode can do through the $W^\pm$ decay 
correlations in $e^+e^-$ collisions using a perfect detector.  

To conclude, (linearly) polarized photons by backscattered laser 
beams of adjustable frequencies at a TeV scale $e^+e^-$ linear 
$e^+e^-$ collider provide us with a very efficient mechanism to 
probe CP violation in two-photon
collisions.
\vskip 0.5cm

\section*{Acknowledgements}

The authors would like to thank F.~Boudjema, F.~Cuypers, I.F.~Ginzburg,
H.S.~Song, R.~Szalapski, T.~Takahashi, C.P.~Yuan, and P.M.~Zerwas for 
useful suggestions and helpful comments.  
The work of SYC is supported in part by the Japan Society for 
the Promotion of Science (No.~P-94024) and that of KH by the 
Grant-in-Aid for Scientific Research from the Japanese Ministry 
of Education, Science and Culture (No.~05228104). The work of 
MSB is supported by Center for Theoretical
Physics, Seoul National University.

\newpage

\begin{center}
{\large FIGURES}
\end{center}

\begin{itemize}
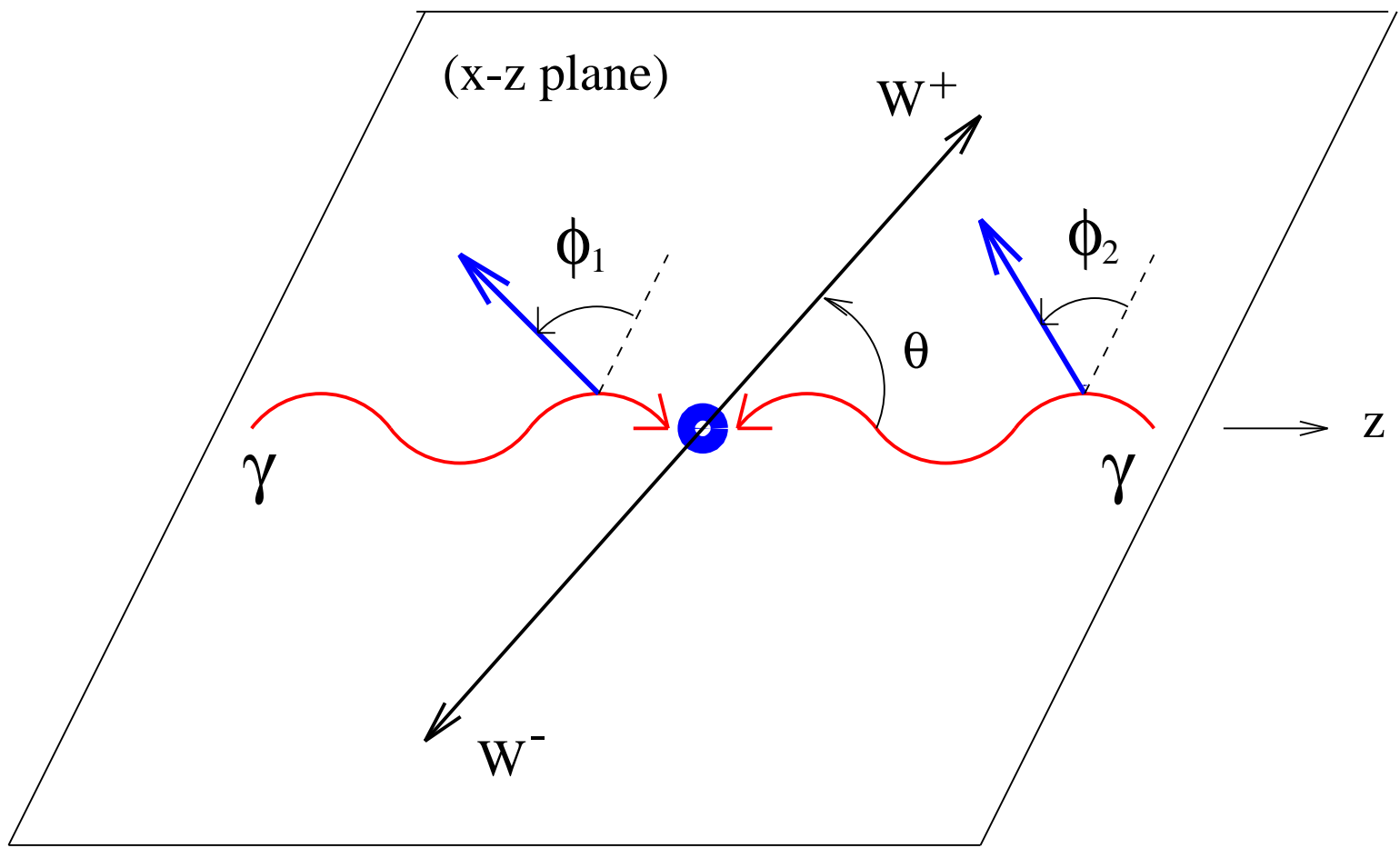
\item[{FIG.~1.}] 
         The coordinate system in the colliding $\gamma\gamma$ 
         c.m. frame. The scattering angle, $\theta$, and the 
         azimuthal angles, $\phi_1$ and $\phi_2$, for the linear
         polarization directions measured from the scattering plane
         are given.
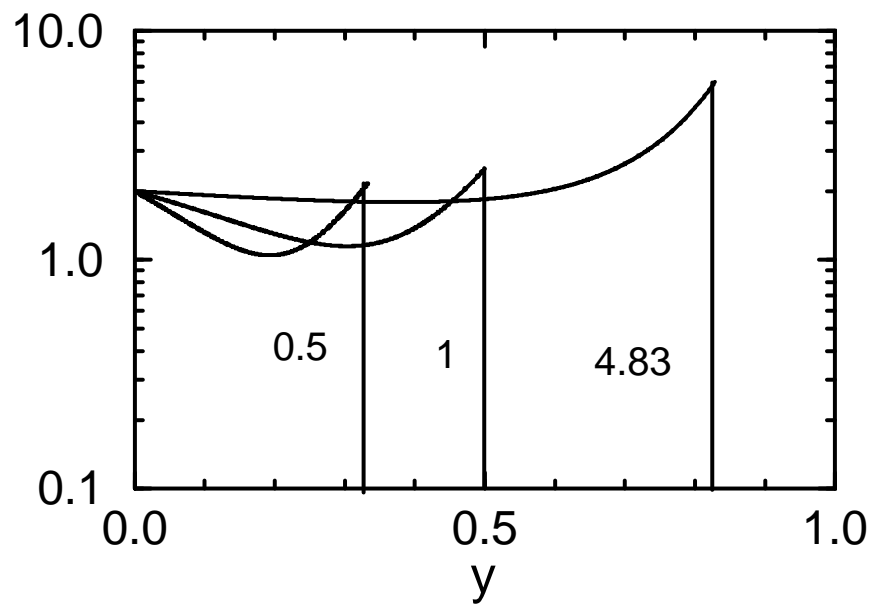
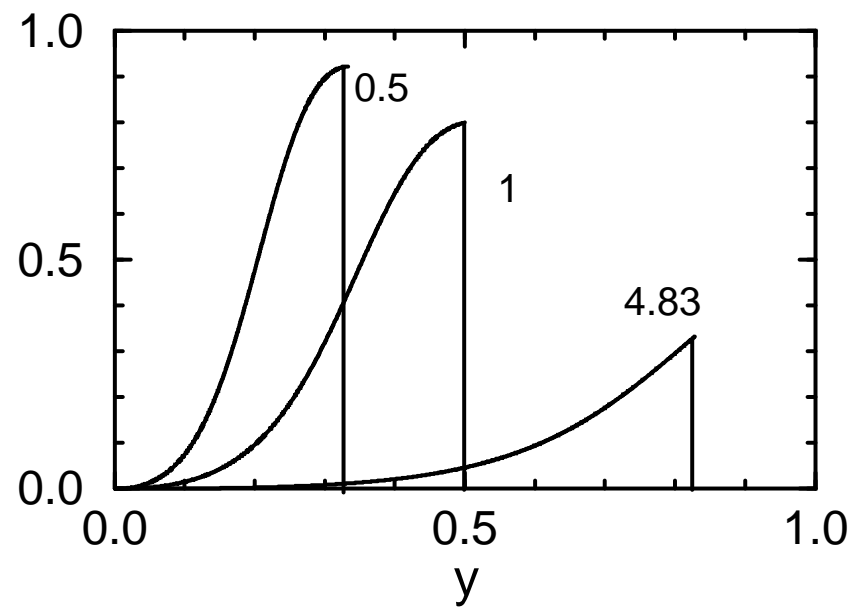
\item[{FIG.~2.}] 
         (a) the photon energy spectrum and (b) the degree of 
         linear polarization of the Compton backscattered photon beam
        for $x=4E\omega_0/m^2_e=0.5$, $1$ and $4.83$.
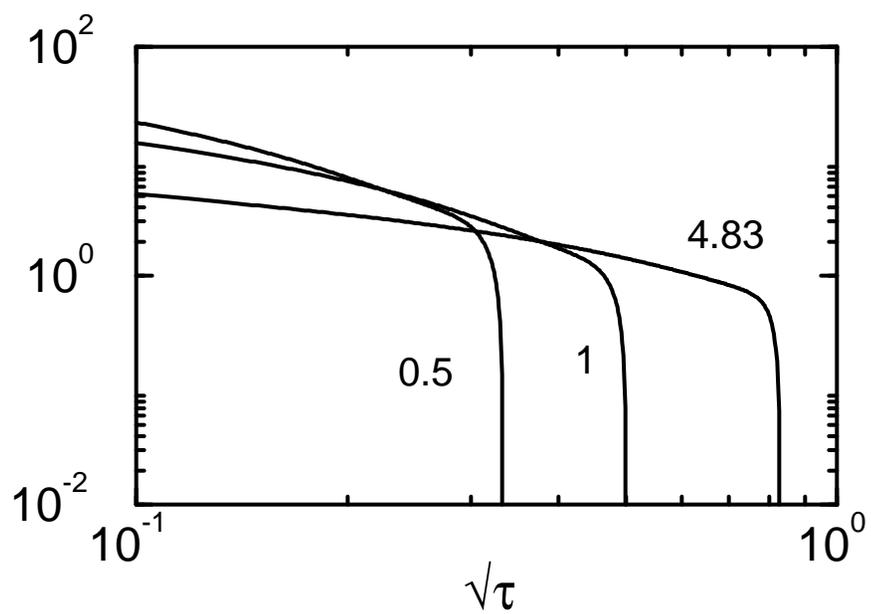
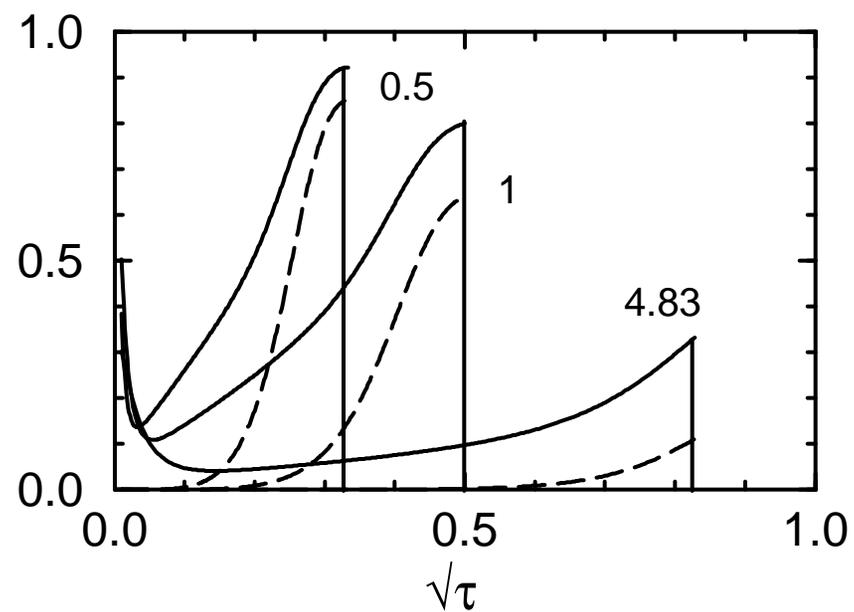
\item[{FIG.~3.}] 
          (a) the $\gamma\gamma$ luminosity spectrum and (b) the
          two linear polarization transfers, $A_\eta$ (solid lines)
          and $A_{\eta\eta}$ (dashed lines), for $x=4E\omega_0/m^2_e=0.5$, 
          $1$ and $4.83$.
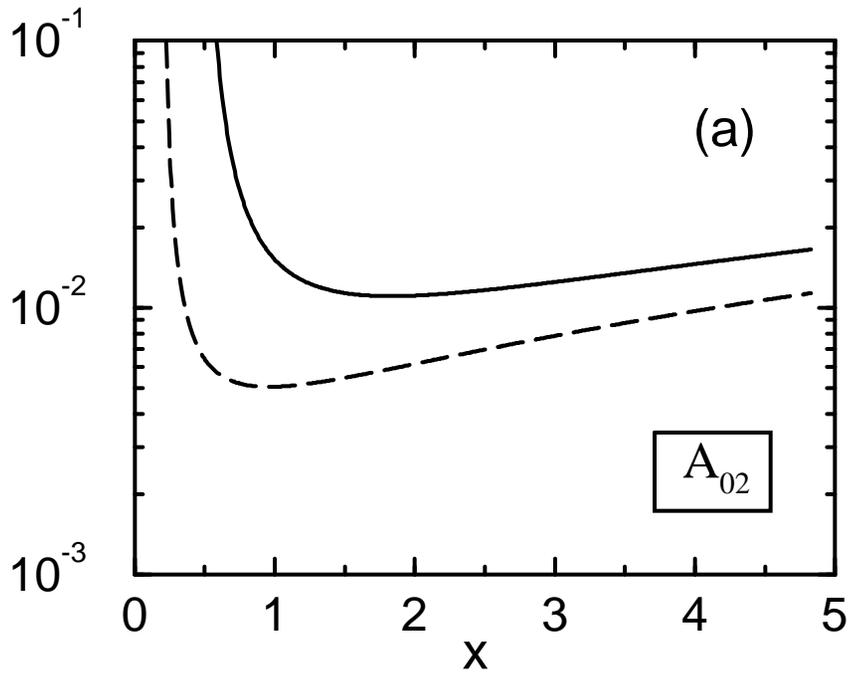
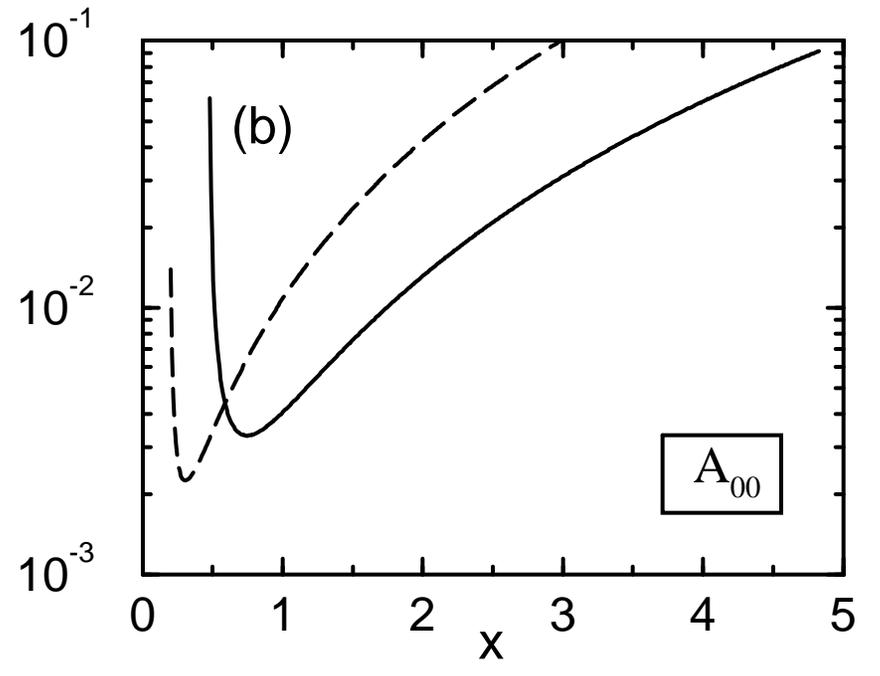
\item[{FIG.~4.}] 
         The $x$ dependence of the $R(Y_1)$ upper bound, 
         ${\rm Max}(|R(Y_1)|)$, at $\sqrt{s}=0.5$ and $1.0$ TeV,
         from (a) the asymmetry $A_{02}$ and (b) the asymmetry
         $A_{00}$, respectively. The solid lines are for
         $\sqrt{s}=0.5$ TeV and the long-dashed lines 
         for $\sqrt{s}=1.0$ TeV. 
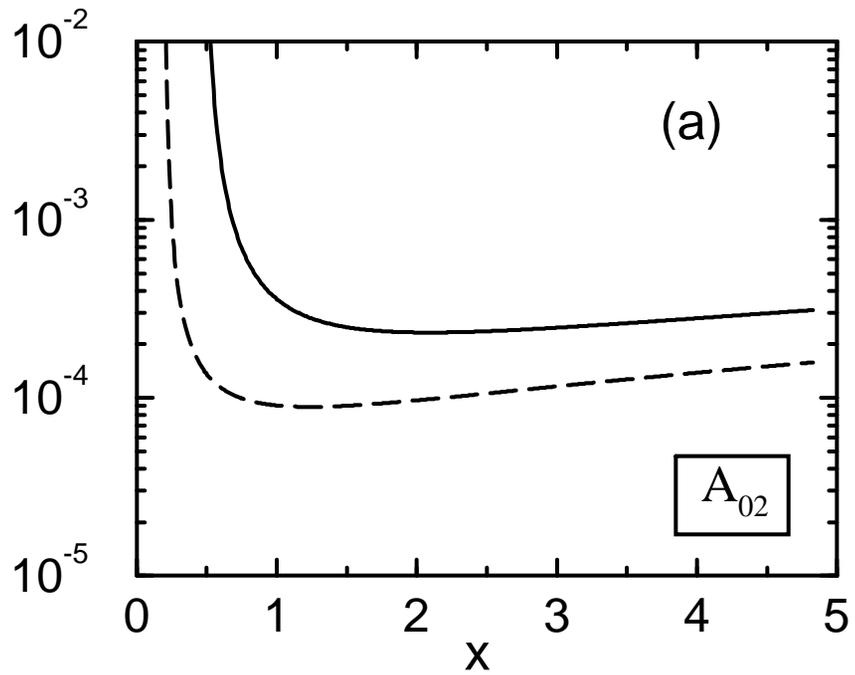
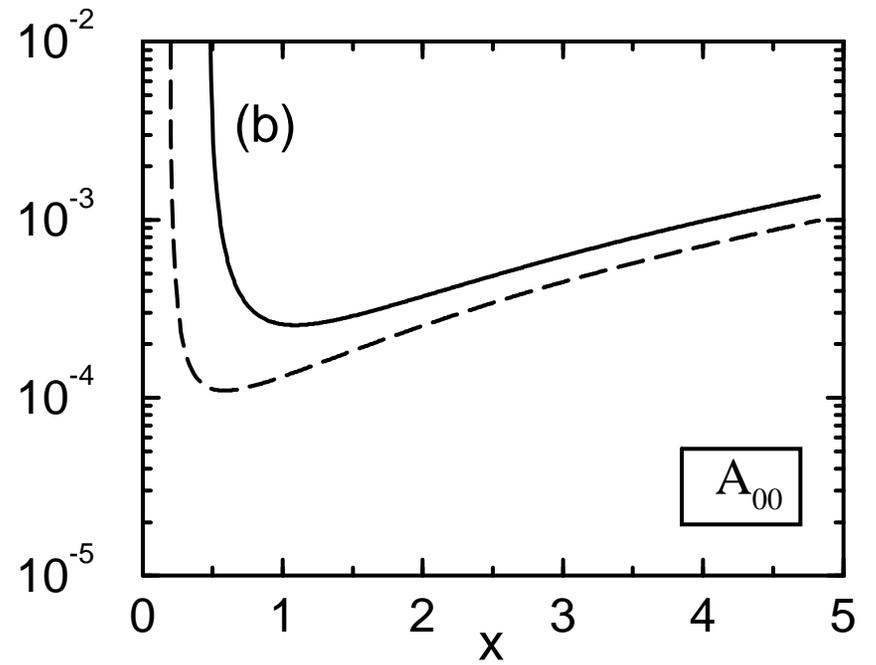
\item[{FIG.~5.}]  
         The $x$ dependence of the $R(Y_2)$ upper bound, 
         ${\rm Max}(|R(Y_2)|)$ at $\sqrt{s}=0.5$ and $1.0$ TeV,
         from (a) the asymmetry $A_{02}$ and (b) the asymmetry
         $A_{00}$, respectively. The solid lines are for
         $\sqrt{s}=0.5$ TeV and the long-dashed lines 
         for $\sqrt{s}=1.0$ TeV. 
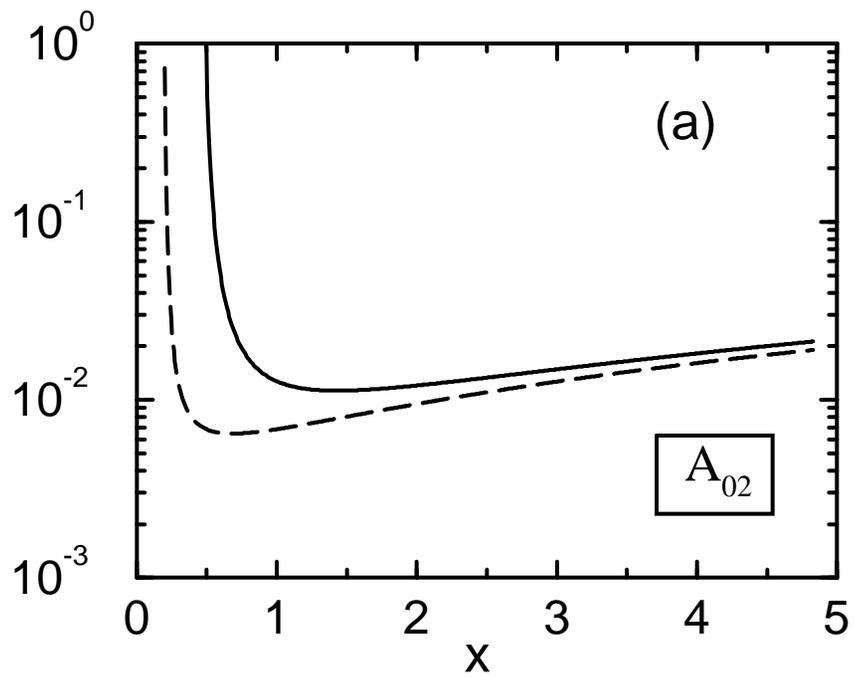
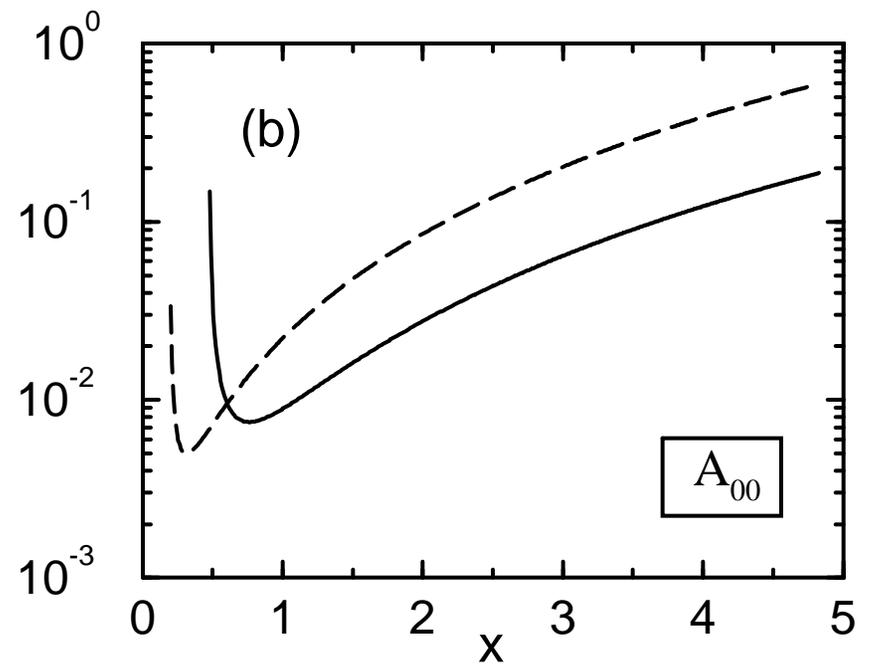
\item[{FIG.~6.}]  
         The $x$ dependence of the $R(Y_4)$ upper bound, 
         ${\rm Max}(|R(Y_4)|)$ at $\sqrt{s}=0.5$ and $1.0$ TeV,
         from (a) the asymmetry $A_{02}$ and (b) the asymmetry
         $A_{00}$, respectively. Here, the Higgs mass is $m_H=100$ GeV.
         The solid lines are for $\sqrt{s}=0.5$ TeV and the long-dashed 
         lines for $\sqrt{s}=1.0$ TeV.
\end{itemize}


\begin{references}
\bibitem{Comment1} The authors in Ref.\ \cite{GbGc} also has 
   considered the 
   distribution ${\cal I}(\Sigma_{02})$, but not the distribution
   ${\cal I}(A_{00})$.
\end{references}
\end{document}